\documentclass[11pt]{article}

\usepackage[english]{babel}

\usepackage[latin1]{inputenc}

\usepackage{amsmath, amsfonts,amssymb}

\usepackage{amstext}

\usepackage{dcpic,pictex}

\usepackage[colorlinks, citecolor=blue, backref]{hyperref}
\usepackage[noblocks]{authblk}

\newtheorem{teorema}{\bf Theorem}

\newtheorem{proposizione}[teorema]{\bf Proposition}

\newtheorem{corollario}[teorema]{\bf Corollary}
\newtheorem{lemma}[teorema]{\bf Lemma}

\newtheorem{definizione}[teorema]{\bf Definition}

\newtheorem{osservazione}[teorema]{\bf Remark}

\newcommand{\Proposizione}[1]{

\addvspace{0.7cm}

\begin{proposizione}{\sf #1}\end{proposizione}}

\newcommand{\Osservazione}[1]{

\addvspace{0.7cm}

\begin{osservazione}{\sf #1}\end{osservazione}}

\newcommand{\Teorema}[1]{

\addvspace{0.7cm}

\begin
{teorema}{\sf #1}\end{teorema}}

\newcommand{\Corollario}[1]{

\addvspace{0.7cm}

\begin{corollario}{\sf #1}\end{corollario}}

\newcommand{\Lemma}[1]{

\addvspace{0.7cm}

\begin{lemma}{\sf #1}\end{lemma}}

\newcommand{\Definizione}[1]{

\addvspace{0.7cm}
\begin{definizione}{\sf #1}\end{definizione}

\addvspace{0.7cm}}

\newcommand{\DIM}[1]{{\bf Proof:} #1 \hspace{\fill}$\clubsuit$

\addvspace{0.7cm}}

\newcommand{\DEF}[1]{\Definizione{#1}}

\newcommand{\THEO}[1]{\Teorema{#1}}

\newcommand{\PROP}[1]{\Proposizione{#1}}
\newcommand{\COR}[1]{\Corollario{#1}}

\newcommand{\REMARK}[1]{\Osservazione{#1}}

\def\id{{\mathchoice
     {\rm 1\kern-0.35em{}1}% Display Math Mode
     {\rm 1\kern-0.35em{}1}% Text Math Mode
     {\rm 1\kern-0.25em{}1}% Subscripts
     {\rm 1\kern-0.25em{}1}% ?
     }}

\def\im{\text{Im}}

\def\be{\begin{equation}}

\def\ee{\end{equation}}

\def\ba{\begin{array}}

\def\ea{\end{array}}

\providecommand{\abs}[1]{\lvert#1\rvert}

\providecommand{\Lie}{\pounds}

\def\<<{``}

\def\R{\mathbb{R}}

\def\N{\mathbb{N}}

\def\D{\partial}

\def\fa{\forall}

\def\map{\longrightarrow}

\def\associa{\longmapsto}

\def\ga{\gamma}

\def\al{\alpha}

\def\B{\beta}

\def\si{\sigma}

\def\Si{\Sigma}

\def\ve{\varepsilon}

\def\la{\lambda}

\def\de{\delta}

\def\om{\omega}

\def\/{\, /\;
}

\begin{document}

\setlength{\parindent}{0in}

\author[a]{Enrico Bibbona}
\author[a,b]{Lorenzo Fatibene}
\author[a,b,c]{Mauro Francaviglia}
\affil[a]{Dipartimento di Matematica, Università degli Studi di Torino (Italy)}
\affil[b]{INFN, Sezione di Torino, Iniziativa Specifica NA12 (Italy)}
\affil[c]{ESG, Università della Calabria (Italy)}
\title{Gauge-natural parameterized variational problems, vakonomic field theories and relativistic hydrodynamics of a charged fluid}
\renewcommand\Authsep{, }
\maketitle

\begin{abstract}
Variational principles for field theories where variations of fields are restricted along a parametrization are considered.
In particular, gauge-natural parametrized variational problems are defined as those in which both the Lagrangian and the parametrization are gauge covariant and some further conditions is satisfied in order to formulate a N\"other theorem that links horizontal and gauge symmetries to the relative conservation laws (generalizing what Fern{\'a}ndez, Garc{\'{\i}}a and Rodrigo did in some recent papers).
The case of vakonomic constraints in field theory is also studied within the framework of parametrized variational problems, defining and comparing two different concepts of criticality of a section, one arising directly from the vakonomic schema, the other making use of an adapted parametrization.
The general theory is then applied to the case of hydrodynamics of a charged fluid coupled with its gravitational and electromagnetic field. A variational formulation including conserved currents and superpotentials is given that turns out to be computationally much easier than the standard one.
\end{abstract}

\section{Introduction}

In some recent literature an interest emerged about variational principles in which some specific requirement of physical or mathematical nature force us to restrict the variations of fields according to some a priori given rule.

\bigskip
As an example, let us mention Mechanics with non-integrable constraints on velocities. As it is well known there are at least two different procedures (so called vakonomic and non-holonomic) which lead to different equations of motion, it is now clear \cite{exp}, at least for affine constraints, that the more realistic trajectories are solutions of the non-holonomic one. However  the situation is less clear in more general cases (see \cite{marle} and references quoted therein), so that we do know very little about constraints in field theory (see \cite{deleon, krupkova}) and we cannot yet decide whether one procedure is better then the other for a reasonable extension out of Mechanics. In any case we shall not enter this question here, but just mention that both procedures can be implemented as variational problems with constrained variations (see \cite{garcia-vak}): they differ exactly in the choice of the rule to be imposed to admissible variations.

\bigskip
Another important case which was recently investigated is Euler-Poincaré reduction (see \cite{epred, reduction, lagred}) where a variational problem with constrained variations arises as a Lagrangian reduction of a free variational problem with symmetries.

\bigskip
A more classical example is presented in the well known book by Hawking and Ellis while introducing the equations of motion for relativistic hydrodynamics \cite{h-e}. There the constraint on variations comes from the physical requirement that variations preserve the continuity equation.

\bigskip
Motivated by all these examples, in this paper we first discuss the formal development of variational problems in which the variations are parametrized by sections of some vector bundle (Section \ref{2}).
This topic was already addressed in \cite{fgr} at least for natural constrained field theories arising from the vakonomic technique; we recover here, and sometimes generalize, some of the results of \cite{fgr} without asking that the parametrization necessarily arises from the vakonomic handling of a constraint, and we develop the case of gauge covariant field theories (much wider then the natural case) for which we develop a full N\"other theory  to link horizontal and gauge symmetries to the relative conservation laws (Section \ref{gau-na}). In Section \ref{vak} we specialize to the case of vakonomic constraints in field theories and define two different concepts of criticality for a section: one arising directly from the vakonomic paradigma, the other following from a compatible parametrization. We also show under what conditions the two concepts coincide.
In Section \ref{es}, finally, our results are applied to the case of relativistic hydrodynamics of a charged fluid.

\bigskip

An advantage of the framework we shall propose hereafter is to provide a unifying language able to deal with all variational problems with constrained variations, no matter whether they come from the vakonomic scheme, from the non-holonomic one, from a reduction problem or from any other physical or mathematical requirement. In our hope, therefore, this unified language can even allow a better comparison between different schemes.

\section{Variational calculus with parametrized variations}
\label{2}
Notation follows \cite{lo}, to which we refer the reader for further details.
Let $C\stackrel{\pi}{\map} M$ be the {\it configuration bundle} of a physical theory the sections $\Gamma(C)$ of which represent the fields (by an abuse of language we will often denote bundles with the same label as their total spaces).

\DEF{A {\it Lagrangian} of order $k$ on the configuration bundle 
$C\stackrel{\pi}{\map}M$ ($m= \dim M$) is a fibered morphism
\[ L: J^k C \map \wedge^m \, T^\ast M.\]}

\DEF{
A {\it parametrization} of order $s$ and rank $l$
($s\geq l$ is required) of the set of constrained variations is a couple
$(E,\mathbb{P})$, where
$E$ is a vector bundle $E\stackrel{\pi_E}{\map} C$ while 
$\mathbb{P}$ is a fibered morphism (section)
\[\mathbb{P}:  J^s C \map (J^l E)^\ast\otimes_{J^lC} VC;\]
here $J$ denotes jet prolongation and $V$ is the vertical functor.
}
If $(x^\mu, y^a,\varepsilon^A)$ are local fibered coordinates on $E$ and $\{\D_a\}$ is the induced fiberwise natural basis of $VC$, a parametrization of order $1$ and rank $1$ associates to a section $y^a(x)$ of $C$ and a section $\varepsilon^A(x, y)$ of $E$ the section \[ \big[p^a_A \big(x^\nu, y^b(x), \D_\nu y^b(x) \big)\varepsilon^A(x,y(x)) + p^{a\phantom{A}\mu}_A (x^\nu, y^b(x), \D_\nu y^b(x) \big) d_\mu \varepsilon^A(x,y(x))\big]\, \D_a\] of $VC$.

\DEF{\label{xxx}Given a compact submanifold $D \subset M$ with suitably regular boundary $\D D$, an {\it
admissible variation} (for a Lagrangian with order $k$ and a parametrization of order $s$ and rank $l$) of a configuration $\si\in
\Gamma(C)$ on $D$ is a smooth $1$-parameter family of sections $\{\si_t\}_{t\in ]-1,1[}\subset
\Gamma (\pi ^{-1} D)$ such that
\begin{enumerate}
\item $\si_0 =\si|_D$
\item $\fa t \in ]-1,1[,\quad j^{k-1}\si_t|_{\D D}= j^{k-1}\si|_{\D D} $
\item \label{333}there exists a section $\ve\in\Gamma(E)$ such that $\frac{d}{d t} \si_t |_{t=0} =<\mathbb{P}\,|\,j^l \ve>\circ j^s \si $ and $j^{k+l-1}(\ve \circ \si)|_{\D D}=0$. 
\end{enumerate} }

\DEF{The set $\{C, L, \mathbb{P}\}$ where $C\stackrel{\pi}{\map}M$ is a 
configuration bundle, $L$ is a Lagrangian on $C$ and $\mathbb{P}$ is a 
parametrization of the set of constrained variations is called a \<<parametrized variational problem''.}

\DEF{We define {\it critical} for the parametrized variational problem $\{C, L, \mathbb{P}\}$ those sections of $C$ such that, for any compact domain $D \subset M$ and for any admissible variation $\{\si_t\}$ defined on $D$ one has 
\[\left.\frac{d}{d t } \int_D L \circ j^k
\si_t \right|_{t =0}=0.\]} 

Accordingly, if we use the trivial parametrization $\mathbb{P}: C\map VC^\ast\otimes_C VC$ that to any $p\in C$ associates the identity of $V_pC$ then the third condition becomes empty and we recover free variational calculus.

For an ordinary variational problem with Lagrangian $L= \mathcal{L}\, ds$, criticality of a section of $C$ is equivalent in local fibered coordinates $(x^\mu, y^a)$ to the fact that for any compact $D\subset M$ and for any $V\in \Gamma(VC)$ such that $j^{k-1} (V\circ\si)|_{\D D}=0$ one has
\[ \int_D\left[\frac{\D \mathcal{L}}{\D y^i} V^i(x,y(x)) +
\frac{\D \mathcal{L}}{\D y^i_\mu}
d_\mu V^i (x,y(x))+
\cdots + \frac{\D \mathcal{L}}{\D y^i_{{\mu_1}\cdots{\mu_k}}}
d_{\mu_1} \cdots d_{\mu_k} V^i (x,y(x))\right]\, ds=0. \]

Explicit calculations (see \cite{lo}) show that the local coordinate expressions given above glue
together giving rise to the global expression
\begin{equation}\label{iiii} \fa D\stackrel{\text{\tiny{cpt}}}{\subset}M,\, \fa V\in VC\,
\text{  s. t. } j^{k-1} (V\circ \si )|_{\D 
D}=0,\quad\int_D<\de L\:|\:
j^k V>
\circ j^k \si =0 \end{equation} where $\de L$ is a global fibered (variational)
morphism
\begin{equation}
\label{deltaL}\de L: J^k C \map (J^k VC)^\ast \otimes_{J^k C} 
\Lambda^m T^\ast M.
\end{equation}

The rigorous definition and a short account on variational morphisms will be
given in Appendix \ref{morfismi}, where the reader can find some non-conventional definitions that we have generalized in order to embrace the case of parametrized variations.

\subsection{$\mathbb{P}$-first variation formula and $\mathbb{P}$-Euler-Lagrange equations}
\label{P}

To define criticality for parametrized variational problems with a $k^{th}$-order Lagrangian and a parametrization of rank $l$, we have to restrict variations to those $V\in VC$ in \eqref{iiii} with $j^{k-1} (V\circ \si )|_{\D 
D}=0$ that can be obtained through the parametrization from a section $\varepsilon$ of $E$ satisfying $j^{l+k-1} (\varepsilon \circ \si)|_{\D D}=0$.
Let us make t explicit on a first order Lagrangian and a parametrization with rank $1$ and order $1$: if  $(x^\mu, y^a, \varepsilon^A)$ are local fibered coordinates on $E$ and $ (p^a_A \varepsilon^A + p^{a\phantom{A}\mu}_A  d_\mu \varepsilon^A)\D_a$ is the local representation of $<\mathbb{P}\,|\,j^1 \varepsilon>\circ j^1 \si$, criticality holds if and only if for any compact $D\subset M$, for any section $\varepsilon$ with coordinate expression $\varepsilon^A(x,y)$ such that both $\varepsilon^A(x,y(x))=0$ and $d_\mu \varepsilon^A(x,y(x))=0$ for all $x\in \D D$, we have
\begin{equation} \int_D\left[\frac{\D \mathcal{L}}{\D y^a} (p^a_A \varepsilon^A + p^{a\phantom{A}\mu}_A  d_\mu \varepsilon^A) +
\frac{\D \mathcal{L}}{\D y^a_\mu}
d_\mu (p^a_A \varepsilon^A + p^{a\phantom{A}\nu}_A  d_\nu \varepsilon^A)\right]\, ds=0.\label{local} \end{equation}

To set up an intrinsic characterization of critical sections possibly given by a set of differential equations let us introduce the following procedure.

Let us take any parametrization $\mathbb{P}$ and think of 
it as a morphism

\[\mathbb{P}': J^s C \times J^l E \map  VC \]
linear in its second argument. Take its $k^{th}$-order (holonomic) 
jet prolongation (see Appendix \ref{morfismi}, Definition \ref{prol})
\[j^k\mathbb{P}': J^{s+k} C \times J^{l+k} E \map  J^kVC \]
and read it as a bundle morphism
\[j^k\mathbb{P}':  J^{s+k} C \map (J^{l+k} E)^\ast\otimes_{J^{s+k}C} J^kVC\]
(for simplicity we still denote it by $j^k\mathbb{P}'$, but the same abuse of notation cannot safely be adopted also for $\mathbb{P}$ and $\mathbb{P}'$ because the jet prolongations of the two objects give completely different results).

Let us consider the variational morphism $\de L: J^k C \map (J^k VC)^\ast
\otimes_{J^kC} \Lambda^m T^\ast M$ introduced in formula 
\eqref{deltaL} and
take its formal contraction (see Definition \ref{contr}) with $j^k\mathbb{P}'$. What we 
get is the
variational $E$-morphism

\[<\de L\,|\,j^k\mathbb{P}'>: J^{k+s} C \map (J^{l+k} E)^\ast
\otimes_{J^{l+k}C} \Lambda^m T^\ast M.\]

Criticality of a section $\si$ can now be recasted in the following global requirement 
\begin{gather*}\displaystyle\fa D\stackrel{\text{\tiny{cpt}}}{\subset}M,\; \fa \varepsilon \in
\Gamma(E)\text{ s. t. }j^{l+k-1} (\varepsilon \circ \si)\big|_{\D D}=0 ,\\ \int_D \Big< <\de
L\,|\,j^k\mathbb{P}'>\,\Big|\, j^{l+k}\varepsilon\Big> \circ 
j^{k+s} \si =0\end{gather*}

This intrinsic procedure leads in the case of first order Lagrangians and first order first rank parameterizations to the local expression \eqref{local}.

Now we can try to turn this requirement into a set of differential equations. The first step is to search for a first variation formula for parametrized variational problems.

Thanks to Theorem \ref{splitting}, for any choice of a fibered connection 
$(\ga,\Gamma)$ on $E\map C\map M$ (see Definition \ref{formalcon} and \ref{fibercon}), there
exists a unique pair
$(\mathbb{E},
\mathbb{F})$ of variational morphisms
\[\begin{aligned}{}&\mathbb{E}(L, \mathbb{P}): J^{2k +l +s}C \map 
E^\ast\otimes_C
\Lambda^m T^\ast M\\
&\mathbb{F}(L, \mathbb{P},\ga): J^{2k +l +s-1}C \map
(J^{l+k-1}E)^\ast\otimes_{J^{l+k-1}C} \Lambda^m T^\ast M \end{aligned}\]
reduced with respect to $(\ga, \Gamma)$ (see Definition \ref{redu}) such that $\fa \varepsilon \in
\Gamma(E)$ one has
\begin{equation}\Big< <\de L\,|\,j^k\mathbb{P}'>\,\Big|\,
j^{l+k}\epsilon\Big>=  <\mathbb{E}\,|\,\varepsilon>+ \text{ Div}
<\mathbb{F}\,|\,j^{l+k-1}\varepsilon>.\label{pardeltaLL}\end{equation}
In the previous equation Div stands for the divergence of a variational morphism (Appendix \ref{morfismi}, Definition \ref{Div}) that coincides when computed on a section with the exterior differential of forms.
\DEF{We call {\it $\mathbb{P}$-Euler-Lagrange morphism} the variational
morphism $\mathbb{E}(L, S,\mathbb{P})$ and {\it $\mathbb{P}$-Poincar\'e -Cartan morphism} the variational morphism $\mathbb{F}(L, S, \mathbb{P}, \ga)$.}

\PROP{A section $\si$ of $C$ is critical for the parametrized variational problem $(C,L,\mathbb{P})$ (with $L$ of order $k$ and $\mathbb{P}$ of order $s$ and rank $l$)  if and only if $<\mathbb{E}\,|\,\varepsilon>\circ j^{2k +l +s}\si =0$.}
\DIM{We have 
\begin{gather*}\displaystyle\fa D\stackrel{\text{\tiny{cpt}}}{\subset}M,\; \fa \varepsilon \in
\Gamma(E)\text{ s. t. }j^{l+k-1} (\varepsilon \circ \si)\big|_{\D D}=0 ,\; \int_D \Big< <\de
L\,|\,j^k\mathbb{P}'>\,\Big|\, j^{l+k}\varepsilon\Big> \circ 
j^{k+s} \si =0\\
\Updownarrow\\
\displaystyle\fa D\stackrel{\text{\tiny{cpt}}}{\subset}M,\; \fa \varepsilon \in
\Gamma(E)\text{ s. t. }j^{l+k-1} (\varepsilon \circ \si)\big|_{\D D}=0 ,\\ \int_D <\mathbb{E}\,|\,\varepsilon> \circ j^{2k +l +s}\si+ \int_D d
\big(<\mathbb{F}\,|\,j^{l+k-1}\varepsilon> \circ 
j^{2k +l +s-1} \si\big) =0\\
\Updownarrow\\
\mathbb{E} \circ j^{2k +l +s}\si=0
\end{gather*}
The first equivalence is ensured by the first variation formula \eqref{pardeltaLL} and the Definition \ref{Div}, while
the last holds in force of the fact that $j^{l+k-1} (\varepsilon \circ \si)\big|_{\D D}=0$ and that $\varepsilon$ and $D$ are arbitrary.}

We stress that the number of derivatives of fields in $\mathbb{E}$ and  $\mathbb{F}$ is not sharp: in fact, due to the particular structure of the morphism $<\de L\,|\,j^k\mathbb{P}'>$, it is actually always lesser then the one expected from the naive issue.
Moreover, the morphism $\mathbb{E}(L, \mathbb{P})$ in the splitting \label{pardeltaL1} is truly
unique (i.e. independent on the connection used),  while
$\mathbb{F}(L, \mathbb{P}, \ga)$ is just one representative in the class of variational morphisms fitting into formula \eqref{pardeltaLL} chosen o that it is reduced with respect to the fibered
connection $(\ga,\Gamma)$; more precisely, the final expression of the morphism $\mathbb{F}(L, \mathbb{P}, \ga)$ depends at most on the spacetime connection $\ga$, while if $l+k\leq 2$ (the most common case) also $\mathbb{F}$ can be defined independently on any part of the fibered connection, so that the splitting algorithm can be performed in coordinates simply by integration by parts on the derivatives of the sections of $E$.

In the case $k=1$, $l=1$ and $s=1$, with the same coordinate representation as in formula \eqref{local}, the morphisms $\mathbb{E}$ and $\mathbb{F}$ are the following:
\begin{gather*}<\mathbb{E}\;|\; \varepsilon> \circ j^3\si = \left\{\left(\frac{\D L}{\D y^a} - d_\nu \frac{\D L}{\D
y^a_\nu}\right) p^a_A  - d_\mu \left[ \left(\frac{\D L}{\D y^a} - d_\nu
\frac{\D L}{\D y^a_\nu}\right){p^a_A}^\mu\right]\right\} \varepsilon^A\;  ds \\
<\mathbb{F}\;|\; j^1 \varepsilon> \circ j^2 \si = \left[ \left(\frac{\D L}{\D y^a} - d_\nu
\frac{\D L}{\D y^a_\nu}\right){p^a_A}^\mu
\varepsilon^A+\frac{\D L}{\D y^a_\mu} (p^a_A \ve^A +
{p^a_A}^\nu d_\nu \varepsilon^A)\right]\;ds\end{gather*}
and they can easily be obtained integrating by parts the integrand of equation \eqref{local}.

\subsection{Symmetries}

The second natural question arising in our investigations is whether the link between symmetries and conserved currents given in the standard N\"other theory is preserved if we vary configurations only along the parametrized variations.
First of all the definition of symmetry for a parametrized variational problem has to be changed with respect to the ordinary one in order to ensure the invariance of the parametrization as well as that of the Lagrangian. 
\DEF{Let $C\stackrel{\pi_C}{\map}M$ be a configuration bundle, $L$ a 
Lagrangian of order $k$ and $(E, \mathbb{P})$ a parametrization  of order $r$ and rank $l$ on the vector bundle $E\stackrel{\pi_E}{\map}C$.

Let $Z:E\map TE$ be a vector field on the bundle $E\map M$ such that $Y= T\pi_E Z$ is a vector field on $C$, while $X=T\pi_E Y$ is a vector field on $M$.

Calling $\Psi_{Z,s}^E$, $\Psi_{Y,s}^C$ and $\Psi_{Y,s}^{VC}$ the flows of fibered transformations generated by the vector fields $Z$ and $Y$ on the bundles $E$, $C$ and $VC$ respectively, we can drag any section $\varepsilon\in \Gamma(E)$ along $\Psi_{Z,s}^E$ getting the section $\varepsilon_s=(\Psi_{Z,s}^E)^\star \varepsilon$, any section $\si\in \Gamma(C)$ along $\Psi_{Y,s}^C$ getting $\si_s=(\Psi_{Y,s}^C)^\star \si$ and, for any section $V\in \Gamma(VC)$, we can also drag $V\circ \si$ along $\Psi_{Y,s}^{VC}$ getting $(\Psi_{Y,s}^{VC})^\star V\circ\si$.

The vector field $Z$ is an {\it infinitesimal Lagrangian symmetry} of the (parametrized) variational problem $\{C,L,\mathbb{P}\}$ if the following {\it covariance identity}
\begin{equation}<\de L\,|\,j^{k} \Lie_Y \si> = \text{Div} (i_X L) +\text{Div}
<\al\,|\,j^{k} Y> \label{covid}
\end{equation}
holds for some morphism $\al : J^{k+h}C \map 
(J^kTC)^\ast \otimes_{J^kC}
\Lambda^{m-1}T^\ast M$,
and moreover we have
\[ <\mathbb{P}\,|\,j^l \varepsilon_s> \circ j^r \si_s = (\Psi_{Y,s}^{VC})^\star (<\mathbb{P}\,|\,j^l \varepsilon> \circ j^r \si). \	\]
\label{simmetrie}}
The definition of infinitesimal symmetry has now been extended to variational problems with parametrized variations. Nevertheless in order to establish a correspondence between symmetries and conserved currents we need some more restrictive hypothesis.
Usually (in the case with free variations) one considers the covariance identity \eqref{covid} and simply splits $\de L$ in its Euler-Lagrange part plus the divergence of the Poincaré-Cartan part: being the first vanishing on shell, one gets the conserved current as the argument of the residual divergence. 
In our framework, on the contrary, field equations arise from the splitting of $<\de L\,|\,j^k\mathbb{P}'>$ so that we would need a rule to associate to any symmetry $Y$ and to any configuration $\si$ a section of the bundle of parameters that in turn is mapped by the parametrization into $\Lie_Y \si$. Without any further hypothesis such a section is not guaranteed to exist.
In thee next Section we will then state all the necessary assumptions which are needed to get a N\"other theorem in the case of gauge-natural symmetries of gauge-natural variational problems in the next section.

\section{Gauge-natural constrained theories}\label{gau-na}
Gauge-natural field theories have been shown to be the most general setting to
describe Lagrangian field theories with gauge symmetries, including gauge
theories, General Relativity in its different formulations, bosons, spinors and also supersymmetries (see \cite{lo-review, lo}).
In the book \cite{lo} it was also described in detail how to cope with
conservation laws linked to pure gauge (vertical) symmetries and 
how to implement
the conservation of energy and momentum as N\"other currents relative 
to an horizontal lift of space-time diffeomorphisms.
The goal of this Section is to extend the previously known results to the case of constrained
variational problems under suitable conditions. The original material on gauge-natural bundles can be found in \cite{eck}; a standard reference with many theoretical improvements is \cite{kolar}; a more applicative approach can be found in \cite{lo}, while a friendly introduction to the subject is also provided in Appendix \ref{GNmaterial}.

\subsection{Morphisms between gauge-natural bundles}

In this sub-Section we recall the fundamental definitions about morphisms between gauge-natural bundles, their push-forward along gauge transformations and their Lie derivatives. Gauge-natural morphisms are also defined.

For any principal automorphism $\Psi= (\phi, \psi)$ of the principal bundle $P\map M$ let us denote by $\mathfrak{B}(\Psi)= (\phi, \hat{\Psi}_\mathfrak{B})$ the gauge-natural lift of $\Psi$ to the gauge natural bundle $\mathfrak{B}(P)$.

\DEF{A bundle morphism between gauge-natural bundles $\mathbb{M}: 
\mathfrak{B}(P) \map
\mathfrak{D}(P)$ projecting onto the identity is said to be {\it 
gauge-natural} if
for any local principal automorphism $\Psi$ of $P$ the following diagram commutes
\nopagebreak
\[
\begindc{\commdiag}[2]
\obj(20,10)[Bd]{$\mathfrak{B}(P)$}
\obj(20,40)[Bu]{$\mathfrak{B}(P)$}
\obj(50,10)[Dd]{$\mathfrak{D}(P)$}
\obj(50,40)[Du]{$\mathfrak{D}(P)$}
\mor{Bu}{Du}{$\mathbb{M}$} 
\mor{Bd}{Dd}{$\mathbb{M}$} 
\mor{Bu}{Bd}{$\mathfrak{B}(\Psi)$} [\atright,\solidarrow]
\mor{Du}{Dd}{$\mathfrak{D}(\Psi)$} 
\enddc\]
}

\DEF{Let $\mathbb{M}: \mathfrak{B}(P) \map\mathfrak{D}(P)$ be a vertical (i.e. projecting onto the identity) morphism between gauge-natural bundles. We define its {\it push-forward} along the local principal automorphism $\Psi= (\phi, \psi)$ to be the unique vertical morphism 
$\Psi^\star\mathbb{M}: \mathfrak{B}(P) \map\mathfrak{D}(P)$ such that for all section $\si \in \Gamma(\mathfrak{B}(P))$ we have
\[\mathfrak{D}(\Psi)^\star(\mathbb{M}\circ \si) = \Psi^\star\mathbb{M} \circ \mathfrak{B}(\Psi)^\star \si.\]}

As a consequence we have the following explicit rule of calculation: 
\[\Psi^\star \mathbb{M}= \mathfrak{D}(\Psi)\circ \mathbb{M}\circ \mathfrak{B}(\Psi^{-1})  =\hat{\Psi}_\mathfrak{D}\circ \mathbb{M}\circ  \hat{\Psi}^{-1}_\mathfrak{B}.\]

\DEF{\label{LLIIEE}Let $\mathbb{M}: \mathfrak{B}(P) \map \mathfrak{D}(P)$ be a vertical morphism between gauge-natural bundles over the same base and $\Psi_s= (\phi_s, \psi_s)$ a $1$-parameter family of local principal automorphisms having $\Xi\in\Gamma(\text{IGA}(P))$ as generator. We define the {\it Lie derivative} of the morphism $\mathbb{M}$ along $\Xi$ to be the morphism $\Lie_\Xi \mathbb{M}: \mathfrak{B}(P) \map V\mathfrak{D}(P)$ that fulfills
\[\Lie_\Xi \mathbb{M} = - \left.\frac{d}{ds} \Psi_s^\star \mathbb{M}\right|_{s=0}. \]
}
From the definition it follows immediately that the expression of the Lie derivative of a morphism is linked to that of the Lie derivative of sections by means of the following rule: for all $\si\in \Gamma(\mathfrak{B}(P))$ 
\[ \Lie_\Xi \mathbb{M} \circ \si = \Lie_\Xi \, (\mathbb{M}\circ \si) - T\mathbb{M} \circ \Lie_\Xi  \si .\]

If we use coordinates $(x^\mu, y^a)$ on $\mathfrak{B}(P)$ and $(x^\mu, z^A)$ on $\mathfrak{D}(P)$, the local expression of $\mathbb{M}$ is $\mathbb{M}:(x^\mu, y^a) \associa (x^\mu, z^A(x^\mu, y^a))$.

Let us call $\hat{\Xi}_{\mathfrak{B}}$ and $\hat{\Xi}_{\mathfrak{D}}$ the gauge-natural lifts of an infinitesimal generator of principal automorphisms $\Xi$ respectively to $\mathfrak{B}(P)$ and $\mathfrak{D}(P)$, and let 
\[\hat{\Xi}_{\mathfrak{B}}= \Xi^\mu \frac{\D}{\D x^\mu} +\hat{\Xi}_{\mathfrak{B}}^a \frac{\D}{\D y^a}  \qquad\text{ and }\qquad \hat{\Xi}_{\mathfrak{D}}= \Xi^\mu \frac{\D}{\D x^\mu} +\hat{\Xi}_{\mathfrak{D}}^A \frac{\D}{\D z^A}\]
be their coordinate expressions.

The Lie derivative of $\mathbb{M}$ has thence local coordinate expression 
\[\Lie_\Xi \mathbb{M}= \left(\hat{\Xi}_{\mathfrak{D}}^A - \frac{\D z^A}{\D y^a}\hat{\Xi}_{\mathfrak{B}}^a \right) \frac{\D}{\D z^A}. \]

\REMARK{\label{LLIEE2}Let $\mathfrak{B}(P)$ and $\mathfrak{D}(P)$ be gauge-natural bundles of order $(r,s)$ and $(j,k)$ respectively. We can think at the Lie derivative as a fibered morphism
\[\Lie\mathbb{M}: \mathfrak{B}(P)\map \big(J^m \text{IGA}(P)\big)^\ast\otimes_{\mathfrak{D}(P)} V\mathfrak{D}(P)\]
with $m=\max\{r,j\}$ such that if $\Xi$ is vertical (i.e. a section of $\textstyle\frac{VP}{G}\hookrightarrow \text{IGA}(P)$), $\Lie_\Xi \mathbb{M}$ depends in fact on the derivatives of $\Xi$ only up to the order $f=\max\{s,k\}$.}

\PROP{A necessary and sufficient condition for a vertical morphism between
gauge-natural bundles $\mathbb{M}: \mathfrak{B}(P) \map \mathfrak{D}(P)$ to be gauge-natural is that for any $1$-parameter family of local automorphisms $\Psi_s$ of $P$ generated by $\Xi$, one has
\[ \Psi_s^\star \mathbb{M}= \mathbb{M}\qquad\text{or}\qquad \Lie_\Xi \mathbb{M} =0.\]
}
\PROP{Any global morphism $\mathbb{M}: \mathfrak{B}(P) \map \mathfrak{D}(P)$ has to be gauge-natural and any local gauge-natural morphism can be extended to a global one.}
\DIM{(Sketch) A trivialization of $P$ induces both a trivialization of $\mathfrak{B}(P)$ and $\mathfrak{D}(P)$ (see \cite{lo}) and gauge-naturality is exactly the same as invariance with respect to changes of trivializations of $P$. }

\subsection{Variationally gauge-natural morphisms}

In this sub-Section we present some technical properties that follow from requiring a local Lagrangian not exactly to be a gauge-natural morphism (that would imply globality) but to satisfy a slightly more relaxed condition that amounts to say that every gauge transformation has to be a Lagrangian symmetry according to definition \ref{simmetrie} if a gauge-natural parametrization is provided.

This generalization is needed in order to deal with the case presented as an example in Section \ref{esempio}.

The Theorems we shall develop here can be applied to ordinary variational calculus, but they have been introduced with the explicit aim of embracing the case of parametrized variations; their consequences will be analyzed in the next sub-Section.

\DEF{\label{var_g_n}Let $\mathfrak{C}(P)\map M$ be a gauge-natural bundle and $F \map \mathfrak{C}(P)$ be a  vector bundle such that the composition $F \map \mathfrak{C}(P) \map M$ is a gauge natural bundle  on $M$ that we call $\mathfrak{F}(P)$.
A local variational morphism
\[\mathbb{M}: J^k \mathfrak{C}(P)  \map (J^h \mathfrak{F}(P))^\ast \otimes_{J^hC} \Lambda^{m-n}T^\ast(M)\]
is said to be {\it variationally gauge-natural} if for any $1$-parameter family of local automorphisms $\Psi_s$ of $P$ generated by $\Xi$ there exists a $1$-parameter family of variational morphisms $\{\al_s\}$
with
\[\al_s: J^{k-1} \mathfrak{C}(P)  \map (J^{h-1} \mathfrak{F}(P))^\ast {\otimes} \Lambda^{m-n-1}T^\ast(M)\]
such that, for every section $X\in \Gamma(\mathfrak{F}(P))$ one has
\[ \Psi_s^\star <\mathbb{M}\:|\:j^h X>= <\mathbb{M}\:|\:\Psi_s^\star j^h X> +  \text{ Div} <\al_s  \:|\: \Psi_s^\star j^{h-1} X>.\]}
 
\REMARK{In terms of Lie derivatives there are many possible equivalent infinitesimal formulations of the definition of variational gauge-naturality. Let us consider the morphism 
\[\mathbb{M}: J^k \mathfrak{C}(P)  \map (J^h \mathfrak{F}(P))^\ast \otimes_{J^hC} \Lambda^{m-n}T^\ast(M)\]
and let us think of it as a morphism
\[\mathbb{M}': J^k \mathfrak{C}(P) \times J^h \mathfrak{F}(P) \map \Lambda^{m-n}T^\ast(M)\]
linear in its second argument. Let us suppose that $\mathfrak{C}(P)$ is gauge-natural of order $(r,s)$ and $\mathfrak{C}(P)$ of order $(j,k)$ and that $m=\min \{r,j\}$, according to Remark \ref{LLIEE2}; then the Lie derivative of $\mathbb{M}'$ with respect to a gauge generator $\Xi$ is a morphism
\[\Lie_\Xi \mathbb{M}': J^k \mathfrak{C}(P) \times J^h \mathfrak{F}(P) \map  \big(J^{m}\text{IGA}(P)\big)^\ast\otimes_{M} V\Lambda^{m-n}T^\ast(M),\]
linear in the second argument, that can also be thought as
\[\Lie_\Xi \mathbb{M}': J^k \mathfrak{C}(P) \map (J^h \mathfrak{F}(P))^\ast \otimes \big(J^{m}\text{IGA}(P)\big)^\ast\otimes V\Lambda^{m-n}T^\ast(M)\]
where the tensor product is on the base $J^h \mathfrak{C}(P)$ (for simplicity we still denote it by $\Lie_\Xi \mathbb{M}'$).
Using the map $\pi_2$ introduced in Remark \ref{pi2}, we can say that variational gauge-naturality of $\mathbb{M}$ is equivalent to the existence of a morphism
$\al: J^{k-1} \mathfrak{C}(P) \map \big(J^h \mathfrak{F}(P)\big)^\ast \otimes  \big(J^{m}\text{IGA}(P)\big)^\ast\otimes V\Lambda^{m-n}T^\ast(M)$ such that

\[\pi_2 < \Lie_\Xi \mathbb{M}'\:|\: j^h X > =  \text{Div } \Big( \pi_2 \Big<<\al\;|\;j^{m-1}\Xi>  \:\Big|\: j^{h-1} X \Big>\Big), \]
and the link between $\al$ and the $1$-parameter family $\{\al_s\}$ of definition \ref{var_g_n} is 

\[- \frac{d}{ds} \left. \al_s  \right|_{s=0}=  <\al\;|\;j^{m-1}\Xi>.\]}

Concrete examples in local coordinates of how to check variational gauge-naturality of a Lagrangian morphism as well as that of an Euler-Lagrange morphism will be given in the proof of Proposition \ref{gndel}.
\bigskip

In our exposition, two technical lemmas precede the main basic properties of variationally gauge-natural morphisms and in particular of variationally gauge-natural Lagrangians in order to simplify their proofs. 
The first lemma is very well known, but we report it here just to display some coordinate expression which will be easily recognized later. To our knowledge the other statements have not yet been proved in the same generality.

\Lemma{\label{lemma1}If $L=\text{Div } \Lambda$ then for every $X\in\Gamma (VC)$ there exists a  morphism $\de \Lambda$ such that $<\de L \:|\: j^k X> = \text{Div} <\de \Lambda \:|\: j^{k-1} X>$. If $\Lambda$ is global, so is $\de \Lambda$.}

\DIM{Let $\Lambda $ be a local variational morphism
\[\Lambda: J^{k-1} C  \map (J^{k-1}VC)^\ast {\otimes}_{J^{k-1}C} \Lambda^{m-1}T^\ast(M)\]
where $M$ is the base of the configuration bundle $C$.
Let us use coordinates $(x^\mu, y^a_{\bar{\al}})$ on $J^k C$ ($\mu$ runs from $1$ to $\text{dim } M$, while $\bar{\al}$ is a multiindex of length $0\leq\abs{\bar{\al}}<k$) and let us express $\Lambda$ in coordinates as $\Lambda = \Lambda^\mu (x^\mu, y^a_{\bar{\al}})\, ds_\mu$ (with $0\leq\abs{\bar{\al}}<k$) while $L= d_\mu \Lambda^\mu (x^\mu, y^a_{\bar{\al}})\,ds$.

The coordinate expression for the morphism $<\de L\,|\, j^k X>$ is the following
\begin{equation} <\de L\,|\, j^k X> = \frac{\D d_\mu \Lambda^\mu}{\D y^a_{\bar{\nu}}}X^a_{\bar{\nu}} \; ds.\label{!}\end{equation}
where the multiindex $\bar{\nu}$ has length $0\leq\abs{\bar{\nu}}\leq k$.

We have
 
\[\frac{\D d_\mu \Lambda^\mu}{\D y^a_{\bar{\nu}}}= d_\mu \frac{\D \Lambda^\mu}{\D y^a_{\bar{\nu}}} + \de^{\bar{\al}+ \bar{1}_\mu}_{\bar{\nu}} \frac{\D \Lambda^\mu}{\D y^a_{\bar{\al}}}, \]
where the multi-index $\bar{\al}$ has length $\abs{\bar{\al}}= \abs{\bar{\nu}} - 1$ and by $\bar{1}_\mu$ we mean a multiindex with $1$ entry in position $\mu$ and zero elsewhere, while for $\abs{\bar{\nu}}=0$ there is no $\bar{\al}$ to fit into the formula so that one has to set $\de^{\bar{\al}+ \bar{1}_\mu}_{\bar{\nu}}=0$, while for $\abs{\bar{\nu}}=k$ we simply have 
\[d_\mu \frac{\D \Lambda^\mu}{\D y^a_{\bar{\nu}}}=0.\]

Substituting now into \eqref{!} we get

\[<\de L\,|\, j^k X> =  \left[ d_\mu \frac{\D \Lambda^\mu}{\D y^a_{\bar{\nu}}} + \de^{\bar{\al}+ \bar{1}_\mu}_{\bar{\nu}} \frac{\D \Lambda^\mu}{\D y^a_{\bar{\al}}} \right]X^a_{\bar{\nu}}\; ds =d_\mu \left(\frac{\D \Lambda^\mu}{\D y^a_{\bar{\al}}}X^a_{\bar{\al}}\right)\; ds.\]

We have found the local expression of the morphism $\de \Lambda$

\[ \de \Lambda = (\de \Lambda)^\mu \,  ds_\mu\qquad\text{ with }\qquad(\de \Lambda)^\mu =\frac{\D \Lambda^\mu}{\D y^a_{\bar{\al}}}X^a_{\bar{\al}}\]
 of which we wanted to prove the existence. It remains to straightforwardly check that if $\Lambda$ is global so is $\de \Lambda$; and that in this case the previous one is an invariant expression that does not depend on the chosen coordinates.}

\Lemma{\label{lemma2}The gauge-natural lift $j^k\Hat{\Xi}_{V\mathfrak{C}}$ of an infinitesimal generator $\Xi\in \Gamma(\text{IGA}(P))$ of principal automorphisms of $P$ to the $k^{th}$-jet prolongation of the vertical bundle $V\mathfrak{C}(P)$ of a gauge-natural bundle $\mathfrak{C}(P)$ can be computed from the lift $j^k \Hat{\Xi}_\mathfrak{C}$ to the same jet prolongation of the base bundle $\mathfrak{C}(P)$ according to the following rule:
let $(x^\mu, y^a_{\bar{\nu}})$ be fibered coordinates on $J^k \mathfrak{C}(P)$ and $(x^\mu, y^b_{\bar{\nu}}, v^b_{\bar{\nu}})$ on $J^k V\mathfrak{C}(P)$ ($\bar{\nu}$ is a multi-index of length $0\leq \abs{\bar{\nu}}\leq k$). If
\[ j^k \Hat{\Xi}_\mathfrak{C}= \Xi^\mu \frac{\D}{\D x^\mu} + (j^k \hat{\Xi}_{\mathfrak{C}})^b_{\bar{\nu}} \frac{\D}{\D y^b_{\bar{\nu}}} \]
is the coordinate expression of $j^k \Hat{\Xi}_\mathfrak{C}$ then the coordinate expression of 
$j^k \Hat{\Xi}_{V\mathfrak{C}}$ is 
\[j^k \Hat{\Xi}_{V\mathfrak{C}}= \Xi^\mu \frac{\D}{\D x^\mu} +  (j^k \hat{\Xi}_{\mathfrak{C}})^b_{\bar{\nu}} \frac{\D}{\D y^b_{\bar{\nu}}} + {(j^k \Hat{\Xi}_{V\mathfrak{C}})}^b_{\bar{\nu}} \frac{\D}{\D v^b_{\bar{\nu}}} \]
with
\[ {(j^k \Hat{\Xi}_{V\mathfrak{C}})}^b_{\bar{\nu}} =  v^a_{\bar{\mu}}  \frac{\D}{\D y^a_{\bar{\mu}}} {(j^k \hat{\Xi}_\mathfrak{C})}^b_{\bar{\nu}} \] where $\bar{\mu}$ is a multi-index of length $0\leq \abs{\bar{\mu}}\leq k$.}

\DIM{If we have a gauge-natural bundle $\mathfrak{B}(P)$ with a system of local coordinates $(x^\mu, p^A)$, given the gauge-natural lift $\Hat{\Xi}_\mathfrak{B}$ of any infinitesimal generator $\Xi$ of principal automorphisms of $P$ with local coordinate representation
\[\Hat{\Xi}_\mathfrak{B}= \Xi^\mu \frac{\D}{\D x^\mu}+ \Hat{\Xi}_\mathfrak{B}^A \frac{\D}{\D p^A},\]
then the lift $\Hat{\Xi}_{V\mathfrak{B}}$ on the vertical bundle of $\mathfrak{B}$, where we use coordinates $(x^\mu, p^A, \varepsilon^A)$, has the following local representation 
\[\Hat{\Xi}_{V\mathfrak{B}}= \Xi^\mu \frac{\D}{\D x^\mu}+ \Hat{\Xi}_\mathfrak{B}^A \frac{\D}{\D p^A} + \varepsilon^B \frac{\D \Hat{\Xi}_\mathfrak{B}^A}{\D p^B} \frac{\D}{\D \varepsilon^A}.\]
Now plug in $\mathfrak{B}(P)= J^k \mathfrak{C}(P)$ and, thanks to the isomorphism $V (J^k \mathfrak{C}(P))\approx J^k (V \mathfrak{C}(P))$, one gets the thesis.}

\PROP{If $L$ is a variationally gauge-natural (local) variational morphism then the gauge-natural morphism $\de L$ introduced in formula \eqref{deltaL} is variationally gauge-natural, too.\label{gndel}}

\DIM{
Let us introduce a fibered coordinate system $(x^\mu, y^a)$ on the gauge-natural configuration bundle $\mathfrak{C}(P)$. On $J^k \mathfrak{C}(P)$ we can naturally induce the system of fibered coordinates $(x^\mu, y^a_{\bar{\nu}})$ where $\bar{\nu}$ is a multi-index of length $0\leq \abs{\bar{\nu}}\leq k$.

For a $k$-th order Lagrangian and for any section $X$ of the vertical bundle $V\mathfrak{C}(P)$ we have

\[<\de L\,|\, j^k X> = \frac{\D L}{\D y^a_{\bar{\nu}}}X^a_{\bar{\nu}}.\]

Let us adopt the same notation of the previous Lemma for the lifts of infinitesimal generators of automorphisms of $P$ to the configuration bundle and its vertical bundle.

Furthermore let $(r,s)$ be the order of $\mathfrak{C}$, so that the order of $J^k\mathfrak{C}$ is at most $(r+k,s+k)$.

The Lagrangian $L:J^k\mathfrak{C}\map \Lambda^m T^\ast M$ is variationally gauge-natural so that  a morphism
$\al: J^{k-1} \mathfrak{C}(P) \map  \big(J^{r+k-1}\text{IGA}(P)\big)^\ast\otimes V\Lambda^{m-n}T^\ast(M)$ exists such that

\[\pi_2  \Lie_\Xi \mathbb{M}' =  \text{Div } \Big( \pi_2 <\al\;|\;j^{m-1}\Xi> \Big)=\text{Div } \Lambda. \]
having called $\Lambda$ the morphism $\Lambda=\pi_2 <\al\;|\;j^{r+k-1}\Xi>$; let $\Lambda^\mu$ be its components with respect to the decomposition $\Lambda= \Lambda^\mu \,ds_\mu $.

We have in coordinates:
\[ \frac{\D L}{\D y^a_{\bar{\nu}}} d_{\bar{\nu}} \Lie_\Xi  y^a = d_\mu (L \Xi^\mu)  +  d_\mu \Lambda^\mu\]
which is equivalent to 

\begin{equation} - \frac{\D L}{\D x^\al} \Xi^\al - \frac{\D L}{\D y^a_{\bar{\nu}}} {(j^k \hat{\Xi}_\mathfrak{C})}^b_{\bar{\nu}} = L d_\mu \Xi^\mu  + d_\mu \Lambda^\mu \label{gnl}\end{equation}
Thanks to Lemma \ref{lemma2} we can notice that if the order of $\mathfrak{C}(P)$ is $(r,s)$ then the order of $V\mathfrak{C}(P)$ is $(r+1,s+1)$.

Gauge-naturality of $\de L: J^k \mathfrak{C}(P) \map (J^k V\mathfrak{C}(P))^\ast \otimes
\Lambda^m T^\ast M $ means that  there exists a morphism $\B: J^{k-1} \mathfrak{C}(P) \map  (J^{k-1} V\mathfrak{C})^\ast \otimes  \big(J^{r+k}\text{IGA}(P)\big)^\ast\otimes V\Lambda^{m-n}T^\ast(M)$ such that

\[\pi_2 < \Lie_\Xi \mathbb{M}'\:|\: j^k X > =  \text{Div } \Big( \pi_2 \Big<<\B\;|\;j^{r+k}\Xi>  \:\Big|\: j^{k-1} X \Big>\Big). \]

Let us then compute $< \Lie_\Xi \mathbb{M}'\:|\: j^h X >$. We have (both multiindices $\bar{\nu}$ and $\bar{\mu}$ have length $0\leq \abs{\bar{\nu}}=\abs{\bar{\mu}}\leq k$).

\begin{gather*}\pi_2\,< \Lie_\Xi \mathbb{M}'\:|\: j^k X > =\left\{
 X^a_{\bar{\mu}} \frac{\D}{\D y^b_{\bar{\nu}}}\frac{\D L}{\D y^a_{\bar{\mu}}} d_{\bar{\nu}} \Lie_\Xi y^b - d_\al \left[\left( \frac{\D L}{\D y^b_{\bar{\nu}}} X^b_{\bar{\nu}} \right) \Xi^\al \right] + \frac{\D L}{\D y^b_{\bar{\nu}}} d_{\bar{\nu}} \Lie_\Xi  X^b \right\} \;ds=\\
=\left\{ - X^a_{\bar{\mu}} \left[ \frac{\D}{\D y^a_{\bar{\mu}}} \frac{\D L}{\D x^\al} \Xi^\al + \frac{\D}{\D y^a_{\bar{\mu}}}\frac{\D L}{\D y^b_{\bar{\nu}}} {(j^k \hat{\Xi}_\mathfrak{C})}^b_{\bar{\nu}}\right] - \frac{\D L}{\D y^b_{\bar{\nu}}} d_\al \left(  X^b_{\bar{\nu}}  \Xi^\al \right) + \frac{\D L}{\D y^b_{\bar{\nu}}} d_{\bar{\nu}} \Lie_\Xi X^b \right\} \;ds. \end{gather*}
Integrating by parts the first two addenda and applying \eqref{gnl} we get

\[\begin{aligned}
\pi_2\,< \Lie_\Xi \mathbb{M}'\:|\: j^k X > =&
\left\{ X^a_{\bar{\mu}}  \frac{\D}{\D y^a_{\bar{\mu}}}\left[ L d_\al \Xi^\al + d_\al \Lambda^\al \right] + \frac{\D L}{\D y^b_{\bar{\nu}}} X^a_{\bar{\mu}}  \frac{\D}{\D y^a_{\bar{\mu}}} {(j^k \hat{\Xi}_\mathfrak{C})}^b_{\bar{\nu}} +\right.\\
&\left.- \frac{\D L}{\D y^b_{\bar{\nu}}} X^b_{\bar{\nu}} d_\al \Xi^\al -  \frac{\D L}{\D y^b_{\bar{\nu}}} \Xi^\al d_\al X^b_{\bar{\nu}} +\frac{\D L}{\D y^b_{\bar{\nu}}}
d_{\bar{\nu}} \Lie_\Xi X^b\right\} \;ds=\\
=&\left\{\frac{\D L}{\D y^b_{\bar{\nu}}} X^a_{\bar{\mu}}  \frac{\D}{\D y^a_{\bar{\mu}}} {(j^k \hat{\Xi}_\mathfrak{C})}^b_{\bar{\nu}} - \frac{\D L}{\D y^b_{\bar{\nu}}}
{(j^k \Hat{\Xi}_{V\mathfrak{C}})}^b_{\bar{\nu}} + X^a_{\bar{\mu}}  \frac{\D}{\D y^a_{\bar{\mu}}}\left(d_\al \Lambda^\al \right) \right\} \;ds.
\end{aligned}\]
Thanks to Lemma \ref{lemma2} the first two addenda cancel, while thanks to Lemma \ref{lemma1} the argument of the derivative in the third addendum is the divergence of the morphism $\de \Lambda$ with local expression 
\[ <\de \Lambda\;|\;j^{k-1} X > = \left( X^a_{\bar{\mu}}  \frac{\D \Lambda^\al}{\D y^a_{\bar{\mu}}}  \right)\; ds_\al. \]
Thus we have
\[ \pi_2 \Big<<\B\;|\;j^{r+k}\Xi>  \:\Big|\: j^{k-1} X \Big>= <\de \Lambda\;|\;j^{k-1} X >\]
and $\de L$ is variationally gauge-natural.}

\THEO{\label{gnboundary}The volume part of a variationally gauge-natural (local) variational morphism is gauge-natural (and thence global).}
\DIM{Let $P\map M$ be a principal bundle, $\mathfrak{C}(P)$ a gauge-natural bundle and $\mathfrak{F}(P)$ a gauge-natural vector bundle. Let $\mathbb{M}$ be a gauge-natural morphism
\[\mathbb{M}: J^k\mathfrak{C}(P)\map (J^h\mathfrak{F}(P))^\ast \otimes_{J^h\mathfrak{C}} \Lambda^m T^\ast M\]
that for all $V\in \Gamma(\mathfrak{F}(P))$ splits into
\[
<\mathbb{M}\:|\:j^h V > = <\mathbb{V}\:|\: V > + \text{ Div}<\mathbb{B}
\:|\:j^{h-1} V >.\]
with
\[ \begin{aligned}
{}&\mathbb{V}\equiv \mathbb{V}(\mathbb{M})\: : J^{h+k} \mathfrak{C}(P)   \map
[\mathfrak{F}(P)]^\ast \otimes_{\mathfrak{C}}\Lambda^{m}T^\ast M\\ &\mathbb{B}\equiv
\mathbb{B}(\mathbb{M},\ga)\:  : J^{h+k-1} \mathfrak{C}(P)   \map (J^{h-1}
\mathfrak{F}(P))^\ast
\otimes_{J^{h-1}\mathfrak{C}}\Lambda^{m-1}T^\ast M.
\end{aligned}\]

For any $1$-parameter family of local automorphism $\Psi_s$ of $P$ and for all $V\in \Gamma(\mathfrak{F}(P))$ we have  
\[ \Psi_s^\star <\mathbb{M}\:|\:j^h V>   = \Psi_s^\star  <\mathbb{V}\:|\: V > + 
\Psi_s^\star \left[\rm{Div}<\mathbb{B}\:|\:j^{h-1} V > \right] \]

As it happens whenever we consider a natural bundle on $M$ as gauge-natural on any $P\map M$ the gauge-natural lift on $\Lambda^{m}T^\ast M$ of any gauge transformation coincides with the natural lift of the diffeomorphism onto which the gauge transformation projects, so that if $\Psi_s=(\phi_s, \psi_s)$ then  $\Lambda^m T^\ast(\Psi_s)= \Lambda^m T^\ast(\phi_s)$.
If computed on a section, the divergence coincides with the exterior differential of forms (see Definition \ref{Div}). Accordingly, as the push-forward along a diffeomorphism commutes with the exterior differential we have the following splitting

\[\Psi_s^\star <\mathbb{M}\:|\:j^h V>  = <\Psi_s^\star \mathbb{V}\:|\:\Psi_s^\star V > + \text{ Div} \left[ \Psi_s^\star <\mathbb{B}\:|\:j^{h-1} V > \right]. \]

By variational gauge-naturality of $\mathbb{M}$ we also have a $1$-parameter family of morphisms $\{\al_s\}$ such that
\begin{align*}
\Psi_s^\star <\mathbb{M}\:|\:j^h V> & = <\mathbb{M}\:|\: \Psi_s^\star j^h V> +  \text{ Div} <\al_s \:|\: \Psi_s^\star (j^{h-1} V)> \\
& = <\mathbb{V}\:|\: \Psi_s^\star  V > +
\text{ Div}<\mathbb{B}+\al_s \:|\:\Psi_s^\star ( j^{h-1}V) >.
\end{align*}

Thanks to the uniqueness of the volume part of the splitting of a variational morphism we can conclude
\[ \Psi_s^\star \mathbb{V} = \mathbb{V} \]
that ensures gauge-naturality of $\mathbb{V}$.}

\subsection{Gauge-natural variational calculus with parametrized variations}

\DEF{\label{problem}A gauge-natural variational problem with parametrized variations is defined by 
the set $(\mathfrak{C}(P),L, \mathfrak{F}, \mathbb{P}, \mathbb{J},\bar{\om})$ of the following objects:
\begin{enumerate}
\item a gauge-natural configuration bundle $\mathfrak{C}(P)\stackrel{\pi}{\map}M$ of order $(r,s)$
with  structure
$G$-bundle $P\stackrel{p}{\map}M$;
\item a variationally gauge-natural (local) Lagrangian morphism $L$ of order $k$;
\item a vector bundle $F\stackrel{\pi_F}{\map}\mathfrak{C}(P)$ such that the 
composite projection
$F\stackrel{\pi\circ\pi_F}{\map}M$ is a gauge-natural bundle $\mathfrak{F}(P)$ of order $(p,f)$ called {\it bundle of parameters};
\item a gauge-natural parameterizing morphism $\mathbb{P}:  J^1 \mathfrak{C}(P) \map (J^q \mathfrak{F}(P))^\ast\otimes_{J^q\mathfrak{C}(P)} V\mathfrak{C}(P)$ ($q\in\{0,1\}$);
\item a gauge-natural morphism $\mathbb{J}: J^{1} \mathfrak{C} \map (J^{r-q}\text{IGA}(P))^\ast \otimes_{J^1 \mathfrak{C}} \mathfrak{F}(P)$  such that we have $\big< \mathbb{P}\;|\; j^q <\mathbb{J}|j^{r-q} \Xi > \big>\circ \:j^1\si  = \Lie_\Xi \si $;
\item a morphism $\bar{\om} :J^{k}\mathfrak{C}(P) \map \mathcal{C}(M)\times_M 
\mathcal{C}(P)$ that
associates to any configuration a couple $(\gamma, \om)$ where $\gamma$ is a linear connection on $M$
and $\om$ is a principal connection on $P$.
\end{enumerate}}

Let us remark that the connection $\ga$ is needed to construct the $\mathbb{P}$-Poincar\'e -Cartan  morphism $\mathbb{F}(L, S, \mathbb{P}, \ga)$ only if $k+l>2$, otherwise this requirement can be dropped, while the connection $\om$ on $P$ will be needed in every case to distinguish horizontal and vertical gauge symmetries (as we will see in a while).  

\THEO{A gauge-natural variational problem with parametrized variations  $(\mathfrak{C}(P),L,\linebreak \mathfrak{F}, \mathbb{P}, \mathbb{J})$ leads to gauge-natural (global) Euler-Lagrange equations.}
\DIM{According to the splitting
\[\Big< <\de L\,|\,j^k \mathbb{P}'>\,\Big|\,
j^{l+k}\epsilon\Big>=  <\mathbb{E}\,|\,\epsilon>+ \text{ Div}
<\mathbb{F}\,|\,j^{l+k-1}\epsilon>,\]
the Euler-Lagrange equations of a gauge-natural variational problem with parametrized variations arise as the boundary part of a variationally gauge-natural variational morphism; in fact $\de L$ is variationally gauge-natural by Property \ref{gndel}, and so is $\mathbb{P}$ by hypothesis (hence also its jet prolongation). Their contraction is thence gauge-natural by definition.  
Thanks to Property \ref{gnboundary} we can conclude that the equations are gauge-natural.}

\bigskip

Let us now remark that according to the fact that the Lagrangian is variationally gauge-natural, every gauge-natural lift on $\mathfrak{C}(P)$ of an infinitesimal generator $\Xi\in\Gamma(\text{IGA}(P))$ of automorphisms of $P$ (or, in a more physical language, every gauge transformation) turns out to be a symmetry of our variational problem. With the help of the other requirements of Definition \ref{problem} we are able to associate to any of these gauge symmetries a gauge invariant \<<N\"other'' conserved current according to the the rule that follows.

\THEO{{\bf-N\"other theorem-} Let us consider a gauge natural variational problem with parametrized variations $(\mathfrak{C}(P),L, \mathfrak{F}, \mathbb{P}, \mathbb{J})$ (orders are as in Definition \ref{problem}) such that
\[\pi_2  \Lie_\Xi  L =  \text{Div } \Big( \pi_2 <\al\;|\;j^{r+k-1}\Xi> \Big)\]
 The unique fibered morphism (momentum  map)
\[\mathcal{E}: J^{2k+l+s-1}\mathfrak{C}(P)  \map J^{r-q-1}\text{IGA}(P)\otimes \Lambda^{n-1}T^\ast M\]
such that for all $\si \in \Gamma(\mathfrak{C}(P))$ and for all $\Xi\in\Gamma(\text{IGA}(P))$ one has 
\begin{multline*} <\mathcal{E}\;|\; j^{r-q-1} \Xi>\circ j^{2k+l+s-1}\si =\\= \Big(<\mathbb{F}\,|\,j^{l+k-1}\mathbb{J}\circ \Xi>- Tp\,\Xi \lrcorner L -  \pi_2 <\al\;|\;j^{r+k-1}\Xi>\Big)\circ j^{2k+l+s-1}\si \end{multline*}
associates to any couple $(\si, \Xi)$ formed by a section $\si\in \Gamma (\mathfrak{C}(P))$ and an infinitesimal generator of principal automorphisms $\Xi\in \Gamma(\text{IGA}(P))$ a current (i. e., a $(n-1)$-form) that is conserved (closed) on-shell.
Moreover the morphism $\mathcal{E}$ is gauge-natural (global).}

\DIM{By gauge-naturality of $L: J^k \mathfrak{C}\map \Lambda^m T^\ast M$ one can find a local morphism $\Lambda: J^{k-1} \mathfrak{C} \map V\Lambda^{m-1} T^\ast M$ such that
\[\pi_2  \Lie_\Xi L = \text{Div}\,( \pi_2 <\al\;|\;j^{r+k-1}\Xi> ).\]

\[\begin{aligned} \pi_2 \Lie_\Xi L &= <\de L \,| \,j^k \Lie_\Xi \si > -\text{ Div}\, (Tp\Xi \lrcorner L)  =\\
&= \Big< <\de L\,|\,j^k \mathbb{P}'>\,\Big|\,
j^{l+k}<\mathbb{J}\;|\; j^{r-q}\Xi>\Big> -\text{ Div}\, (Tp\Xi \lrcorner L)=\\
&=\Big<\mathbb{E}\,\Big|\,<\mathbb{J}\;|\; j^{r-q}\Xi>\Big>+ \text{ Div}\Big<\mathbb{F}\,\Big|\,j^{l+k-1}<\mathbb{J}\;|\; j^{r-q}\Xi>\Big>-\text{ Div}\, (Tp\Xi \lrcorner L). \end{aligned}\]
Thence
\begin{multline*}- \Big<\mathbb{E}\,\Big|\,<\mathbb{J}\;|\; j^{r-q}\Xi>\Big> =\\= \text{ Div} \Big<\mathbb{F}\,\Big|\,j^{l+k-1}<\mathbb{J}\;|\; j^{r-q}\Xi> \Big> -\text{ Div}\, (Tp\Xi \lrcorner L) - \text{Div}\, (\pi_2 <\al\;|\;j^{r+k-1}\Xi> )\end{multline*}
so
\begin{equation} - \Big<\mathbb{E}\,\Big|\,<\mathbb{J}\;|\; j^{r-q}\Xi>\Big> = \text{ Div} <\mathcal{E}\;|\; j^{r-q-1} \Xi>. \label{KKKJ}\end{equation}
While from the definition of $\mathcal{E}$ one can hardly argue how many derivatives of $\Xi$ will appear in its final expression (cancellations may arise!) from the latter  identity (that holds off-shell) we can say that $\Xi$ can enter with its derivatives up to the order $r-q-1$.
If $\si$ is a solution of the corresponding Euler-Lagrange equations then we have
\[d \left[\Big(\,\Big<\mathbb{F}\,\Big|\,j^{l+k-1}<\mathbb{J}\;|\; j^{r-q}\Xi>\Big>- Tp\Xi \lrcorner L - \pi_2 <\al\;|\;j^{r+k-1}\Xi> \Big)\circ\,j^{2k+l+s-1}\si\right]=0.\]
To prove gauge-naturality of $\mathcal{E}$ let us notice that identity \eqref{KKKJ} holds; thence, as the push-forward and the exterior differential of forms do commute, it is sufficient to prove gauge-naturality of the left hand side.
The morphism $\mathbb{E}$ is gauge-natural by Property \ref{gnboundary}, and it is contracted on $\mathbb{J}$ that is gauge-natural by hypothesis; thus the thesis follows.}

In the natural case we were able to construct N\"other currents related to diffeomorphisms, and their integrals on $(m-1)$-surfaces were linked to the energy-momentum of the system.
In the gauge-natural case we cannot any longer lift diffeomorphisms of $M$, but according to \cite{lo}, if on $P$ we have a connection gauge-naturally derived from fields, for any infinitesimal generator of principal automorphisms we can define the horizontal part with respect to this connection, and compute the conserved N\"other currents relative to it. This should be regarded as the density of energy momentum in a gauge-natural variational principle also in presence of a parametrization relative to a constraint.

\section{Constrained variational calculus and vakonomic field theories}
\label{vak}
Let us now apply the previously developed formalism to field theories with constraints within the vakonomic approach.
The non-holonomic technique can be studied within the same formalism too, but with a different characterization of admissible variations; this case will be considered in a forthcoming paper.

\subsection{Vak-criticality} \label{1}
The definitions we will use to implement the principle of constrained 
least action
are a generalization of the ones given by Arnold and others in 
\cite{arnold-vak}
in the case of Vakonomic Mechanics. We will abandon their line when they
introduce Lagrange multipliers to characterize critical sections.

\DEF{Let $C\stackrel{\pi}{\map} M$ be the configuration bundle. A {\it constraint} of order $p$ and codimension $r$ is a submanifold $S\subset J^p C$ of codimension $r$ that by
$\pi^p_{p-1}$ projects onto the whole $J^{p-1} C$.}

\DEF{A configuration $\si\in\Gamma(C)$ is said to be {\it admissible} 
with respect
to $S$ if its $p^{th}$-order jet prolongation lies in S. The space of admissible
configuration with respect to $S$ is
\[ \Gamma_S(C)= \{\si\in\Gamma(C)\/ \im (j^p\si) \in S  \}.\] }

\DEF{Given a compact submanifold $D\stackrel{cpt}{\subset} M$, a {\it vak-admissible variation} (at order $k$) of an admissible configuration $\si\in
\Gamma_S(C)$ is a smooth $1$-parameter family of sections $\{\si_\ve\}_{\ve\in ]-1,1[}\subset
\Gamma (\pi ^{-1} D)$ such that
\begin{enumerate}
\item $\si_0 =\si|_D$
\item $\fa \ve \in ]-1,1[,\quad j^{k-1}\si_\ve|_{\D D}= j^{k-1}\si|_{\D D} $
\item $\frac{d}{d \ve}j^p \si_\ve |_{\ve=0} \in J^p VC \cap TS$. \label{tre}
\end{enumerate} }

\DEF{The set $\{C, L, S\}$ where $C\stackrel{\pi}{\map}M$ is a 
configuration bundle, $L$ is a Lagrangian on $C$ and $S$ is a 
constraint is called a \<<constrained variational problem''.}

\DEF{We say that an admissible section $\si\in \Gamma_S(C)$ is {\it 
vakonomically critical} (or vak-critical) for the variational problem $\{C, L, S\}$ if
$\fa D\stackrel{\text{\tiny{cpt}}}{\subset}M$ for any vak-admissible variation
$\{\si_\ve\}_{\ve\in ]-1,1[}\in
\Gamma (\pi ^{-1} D)$ we have
\[\left.\frac{d}{d \ve } \int_D L \circ j^k
\si_\ve \right|_{\ve =0}=0. \] }

An equivalent infinitesimal condition is that \[\fa D\stackrel{\text{\tiny{cpt}}}{\subset}M,\, \fa V\in VC\,
\text{  s. t. } j^p V \in TS\text{ and }j^{k-1} V|_{\D 
D}=0,\quad\int_D<\de L\:|\:
j^k V>
\circ j^k \si =0 \] where $\de L$ is the fibered (variational)
morphism defined in \eqref{deltaL}.

\subsection{Parametrized variations and vakonomic criticality}
In order to turn the vak-criticality condition into a differential equation, the classical strategy (at least in Mechanics) is to introduce Lagrange multipliers (see \cite{arnold-vak}).
Here we try to profitably use a suitable parametrization of the set of constrained variations. This idea was first formally developed by Fernandez, Garcia and Rodrigo in the very interesting paper \cite{fgr}. 

\DEF{Let $C\stackrel{\pi}{\map}M$ be the configuration bundle.
A {\it parametrization}
\[\mathbb{P}_S:  J^s C \map (J^l E)^\ast\otimes_{J^sC} VC\]
of order $s$ and rank $l$
($s\geq l$ is required) of the set of constrained variations is said to be {\it vakonomically adapted} to the constraint $S\subset J^pC$ if for all $ \varepsilon\in\Gamma(E)$ and $\si\in \Gamma_S(C)$
the vertical vector field
$j^p (<\mathbb{P}\,|\,j^l \varepsilon>\circ j^s \si)\in TS$.
}

\DEF{A parametrization $\mathbb{P}_S:J^s C \map (J^l E)^\ast\otimes_{J^sC} VC$
vak-adapted to a constraint $S$ is said to be {\it faithful on $\si\in \Gamma_S(C)$ to $S$} if for all $V\in VC \text{ such 
that both } j^p (V \circ\si) \in
TS$ and $j^{k-1} (V \circ\si)|_{\D D}=0$ hold, there exist a section 
$\varepsilon\in\Gamma(E)$ such that $<\mathbb{P}\,|\,j^l
\varepsilon>\circ j^s \si = V$ and $j^{k-1+l} (\varepsilon\circ\si)|_{\D D}=0$}

The fundamental problem of the existence of a (possibly faithful) parametrization vakonomically adapted to a constraint $S$ has not yet been studied in general and will not be faced here. For the moment we
limit ourselves to consider a given specific parametrization as
a part of the variational problem, deferring the study of the general case to
further investigations. The same was done in all the previous papers 
on this topic (see e.g. \cite{fgr}).

\bigskip

Once we have a vak-adapted parametrization $\mathbb{P}_S$ we can study $\mathbb{P}_S$-criticality, constructing, as we have shown in Section \ref{P}, the relevant $\mathbb{P}_S$-Euler-Lagrange equations.

It is important to remark that even having defined the Lagrangian 
on the whole configuration bundle, the first variation formula arising from it does not
depend on the value of the Lagrangian outside the constraint. In fact 
if one has a
Lagrangian $ \hat{L}: J^k C \map \wedge^m \, T^\ast M$ such that
$\hat{L} = L +N$ with $N$ proportional to the equation of the constraint $S$, then on the constraint $S$ it automatically holds $<\de N \:|\: 
j^k V>\circ
j^k \si =0$ for all
$V\in VC\,
\text{  s.t. } j^p V \in TS$ and for all $\si \in \Gamma_S(C)$ therefore the
addendum $N$ does not contribute to the first variation formula, 
neither in the field
equation nor in the boundary part.

The link between vak-criticality and $\mathbb{P}_S$-criticality is given by the following Proposition.

\PROP{\label{mot}Let $\si\in \Gamma_S(C)$ be a vakonomically critical section 
for the constrained variational problem
$\{C, L, S\}$; then for any adapted parametrization $\mathbb{P}_S$ $\si$ is $\mathbb{P}_S$-critical.}
\DIM{We have
\begin{gather*}
\si\in \Gamma_S(C) \text{ is  critical}\\
\Updownarrow \\
\fa D\stackrel{\text{\tiny{cpt}}}{\subset}M,\; \fa \text{ adm. var. }
\{\si_\ve\},\quad\left.\displaystyle\frac{d}{d \ve } \int_D L \circ
j^k
\si_\ve
\right|_{\ve =0}=0\\
\Updownarrow \\
\fa D\stackrel{\text{\tiny{cpt}}}{\subset}M,\, \fa V\in VC\,
\text{  s. t. } j^p V \in TS\text{ and }j^{k-1}\si|_{\D D}=0,\quad\int_D<\de L\:|\:
j^k V>
\circ j^k \si =0\\
\phantom{(a)}\Downarrow \text{(a)}\\
\displaystyle\fa D\stackrel{\text{\tiny{cpt}}}{\subset}M,\; \fa \varepsilon \in
\Gamma(E)\text{ s. t. }j^{l+k-1}\varepsilon|_{\D D}=0 ,\quad\int_D \Big< <\de
L\,|\,j^k \mathbb{P}'>\,\Big|\, j^{l+k}\varepsilon\Big> \circ 
j^{k+s} \si =0\\
\Updownarrow \\
\mathbb{E}\circ j^{2k+l+s}\si=0.
\end{gather*}
Step (a) is not an equivalence because there can be admissible 
infinitesimal variations
vanishing at the boundary with their derivatives up to the desired 
order that do not come
from sections of the bundle of parameters that do vanish on the 
boundary. The last
equivalence holds in force of Stokes's theorem, the vanishing of
$j^{l+k-1}\varepsilon$ on the boundary and
the independence of the generators of $E$. }
\COR{Let $\si\in \Gamma_S(C)$ be an admissible $\mathbb{P}_S$-critical section for 
the constrained variational problem $\{C, L, S\}$ and let 
also $\mathbb{P}_S$ be
faithful to $S$ on $\si$; then $\si$ is vakonomically critical for the 
constrained variational
problem $\{C, L, S\}$.}

\section{Examples: }
\label{es}
To our knowledge, parametrized variational problems were first formally investigated by {Fer-n\'andez}, Garc\'ia and Rodrigo \cite{fgr} in relation with their application to Lagrangian reduction and in particular to Euler-Poincar\'e reduction \cite{epred, lagred, reduction}.
They were already implicitly used in the literature to deal with Vakonomic Mechanics \cite{arnold-vak, garcia-vak} and Relativistic Hydrodynamics \cite{h-e}.
What we want to present here is a gauge-natural example that also helps to clarify the distinction between a vak-critical section of a constrained variational problem and a $\mathbb{P}_S$-critical section, of which we study the equations of motions and the conserved N\"other currents. 

\subsection{A gauge-natural example: charged general relativistic fluids.}
\label{esempio}
A charged general relativistic fluid can be thought as a congruence 
of curves filling  a
causal domain $D\subseteq M$, each of them carrying some rest mass and 
charge, being source of gravity and of electromagnetic interaction.

Our kinematical description is inspired by \cite{h-e}, 
where the gravitational
field is represented by a Lorentzian metric with components $g_{\mu\nu}$, the 
electromagnetic potential by a
connection $A_\mu$ on a principal $U(1)$-bundle $P$ and the fluid 
degrees of freedom are represented by a nowhere vanishing $3$-form $J=J^\mu ds_\mu$ such that 
the unit  timelike vector
$u^\mu=J^\mu /\abs{J}$ is tangent to the flow lines and that 
the volume form
$\sqrt{\abs{g}}\rho=\abs{J}$ represents the matter density in D.
For what concerns the charge density we simply assume that at each spacetime 
point it is proportional  to
the matter density by an \<<elementary charge'' $q$.

An alternative (more common) approach to describe the fluid degrees of freedom (see \cite{taub, kijowski-art, brown}) is to associate to any space-time point $x^\mu$ three scalar fields $R^a(x^\mu)$ $(a=1,2,3)$ identifying the \<<abstract fluid particles''.   
We remark that using the tangent vector to flow lines as 
fluid variable, we loose the direct
kinematical identification between different points of the same flow line. In fact, having a
configuration $J^\mu (x)$, in order to know whether two points of spacetime are connected by the same flow line, we have to
solve the ODE that defines the integral curves of $u^\mu$ with 
initial condition in one of them.

In order to implement conservation of matter we only 
allow for closed $J$, so that for any
closed  $3$-surface $\Si$ in $D$ the flow of $J$ through $\Si$ is zero.
The bundle of configurations is thence
\[ \mathfrak{C}(P) = Lor(M) \times_M \mathcal{C}(P) \times_M \Lambda^3 T^\ast M\] 
that is gauge-natural having $P\map M$ with fiber $U(1)$ as structure bundle.
The field $J$, moreover, is subject to the constraint $S\subset j^1 \Lambda^3 T^\ast M$
represented by the equation $dJ=0$, or locally $d_\mu J^\mu=0$ (we also require $J$ to be time-like with respect to the unknown metric, but this just selects an open set in the configuration bundle, so that the condition is automatically preserved by any infinitesimal variation).

\bigskip

The dynamics of the system is governed by the Lagrangian

\begin{equation}
L = L_{H} + L_{EM}+  L_F + L_{int} \label{L1}\end{equation} with

\[\begin{aligned}
{}&L_H ( g_{\al\B}, \D_\la g_{\al\B}, \D_{\la\omega} g_{\al\B}) = \frac{\sqrt{\abs{g}}}{2\kappa}\: R \, ds\\
& L_{EM} (g_{\al\B}, A_\nu, \D_\si A_\nu) = -\frac{1}{4}\sqrt{\abs{g}}\: F^{\rho\si} F_{\rho\si}\, ds\\
& L_F (J^\mu, g_{\al\B}) = -\sqrt{\abs{g}}\: [\rho\;(1+ e (\rho))]\, ds\\
& L_{int} = q J^\mu A_\mu ds
\end{aligned}\]
where $R$ is the scalar curvature of the metric, $F_{\rho\si}$ are 
the curvature coefficients (field strength) of the connection $A$ (gauge potential) and $e (\rho)$ is a 
generic function of the scalar
$\rho= \sqrt{\frac{g_{\mu\nu}J^\mu J^\nu}{\abs{g}}}$, physically 
interpreted as the internal energy of the fluid.
It gives rise to the pressure $P= \rho^2 \frac{\D e}{\D \rho}$.

The gauge-natural bundle of parameters we introduce is then
\[ \mathfrak{F}(P)= (\pi_{\mathfrak{C}(P)})^\ast V Lor(M) \times_{\mathfrak{C}(P)} (\pi_{\mathfrak{C}(P)})^\ast V\mathcal{C(P)} 
\times_{\mathfrak{C}(P)} (\pi_{\mathfrak{C}(P)})^\ast TM\]
and the parametrization is such that for any section 
$\varepsilon=(\de g^{\mu\nu}, \de A_\mu,
X^\mu)
\in
\Gamma (E)$ and for any configuration $\si \in \Gamma (C)$ we have
\[<\mathbb{P}_S\,|\, j^1 \varepsilon>  = \big(\de g_{\mu\nu}(x^\si, 
g_{\al\B}), \de
A_\mu (x,A_\al),  \Lie_X J^\mu \big)  \]
where $\Lie_X J^\mu$ denotes the formal Lie derivative of $J$, i.e. 
$\Lie_X J^\mu = \nabla_\nu
J^\mu X^\nu - J^\nu \nabla_\nu X^\mu + J^\mu\nabla_\nu X^\nu$.
In this way we are freely varying the gravitational and 
electromagnetic degrees of freedom, while
variations of the fluid variables are taken to be tangent to the 
constraint $S$; in fact for every vector field $X$ and for every admissible configuration $J$ (i.e. such that $\D_\mu J^\mu=0$ holds) we have $\D_\mu \Lie_X J^\mu=0$.

Let us remark that this parametrization is trivially gauge-natural; in fact it partially coincides with the identity and partially with the Lie derivative.

\REMARK{
We want to spend some words to stress that our parametrization is adapted to the constraint $S$, but in general not faithful to it; nevertheless this is exactly what Physics requires.
Varying $J$ according to our prescription is equivalent to drag $J$ along a $1$-parameter family of diffeomorphisms generated by $X$ and this amounts to drag the integral curves of $u^\mu$  and to adjust the density $\rho$ to keep $J$ conserved (see \cite{h-e}).
If we consider a compact subset $D\subset M$ and find a tangent vector $X$ whose flow moves the integral curves of $u^\mu$ also on the boundary of $D$ ($X|_{\D D}\neq 0$) but leaves their tangent vectors fixed ($\Lie_X J^\mu|_{\D D}=0$, as it is possible in some cases), we have that it produces a family of deformed $J$ that are fixed on the boundary and that fulfill conservation of matter, but do not come from a vector field that vanishes on the boundary as it would be requested by faithfulness.
Nevertheless, this is exactly what we want to do from the physical viewpoint: when we think of the congruence of curves that represent our fluid and we imagine to vary them without moving the boundary we do not want to move particles, not only tangent vectors.
To support our choice to vary fields along our parametrization, we stress that also in the alternative approach of \<<abstract fluid particles'' variations leave unchanged the particle identification on the boundary.
Solutions of the Euler-Lagrange equations that we are going to find will thence be $\mathbb{P}_S$-critical sections of the parametrized constrained variational problem, without being necessarily vak-critical solutions of the variational problem with the constraint $S$ given by $dJ=0$.}

The first variation formulae arising from the freely varied 
Lagrangians $L_{H}$ and
$L_{EM}$ are the well known

\begin{equation}\label{H}\big< \de L_{H}\,\big| j^2 <\mathbb{P}_S\,|\, j^1 \varepsilon>\big> = 
\frac{\sqrt{\abs{g}}}{2\kappa}
(R_{\mu\nu} -
\frac{1}{2}R g_{\mu\nu})\de g^{\mu\nu}\, ds + \nabla_\al \big< 
\mathbb{F}^\al_{(H)}\,\big|\, j^1
<\mathbb{P}_S\,|\, j^1 \varepsilon> \big>\end{equation} with
\[\big< \mathbb{F}^\al_{(H)}\,\big|\, j^1
<\mathbb{P}_S\,|\, j^1 \varepsilon> \big> =  \frac{\sqrt{\abs{g}}}{2\kappa}
(g^{\al\la}g_{\rho\si}-\de^\la_{(\rho} \de^\al_{\si)}) \nabla_\la \de 
g^{\rho\si}\,ds\]

where the covariant derivative is relative to the Levi-Civita connection and

\begin{equation}\begin{split}\label{EM}\big< \de L_{EM}\,\big| j^1 <\mathbb{P}_S\,|\, j^1 \varepsilon>\big> 
=& - \sqrt{\abs{g}} (\nabla_\mu
F^{\mu\nu}) \de A_\nu \, ds -   \sqrt{\abs{g}} \nabla_\mu (F^{\mu\nu} \de A_\nu)\,ds\\
&-\frac{\sqrt{\abs{g}}}{2}H^{(EM)}_{\mu\nu}\de g^{\mu\nu}\,ds\end{split}\end{equation} with
\[\sqrt{\abs{g}}\, H^{(EM)}_{\mu\nu}\,ds=-2\frac{\D L_{EM}}{\D g^{\mu\nu}}=  \sqrt{\abs{g}}\,
(F_{\mu\rho}F_{\nu\cdot}^{\phantom{\nu}\rho}-\frac{1}{4}
F_{\al\B}F^{\al\B}g_{\mu\nu})\, ds.\]
 From the parametrized variation of $L_{F}$, by taking into account the definitions
\[\mu = \rho\;(1+ e (\rho)) \hbox{\qquad and \qquad}
P = \rho^2 \frac{\D e}{\D \rho}\]
together with their consequences
\[ \rho \, \frac{\D \mu}{\D \rho} = \mu +P
\hbox{\qquad and \qquad} 
\rho \, \nabla_\nu \left( \frac{\D \mu}{\D \rho} \right) = \nabla_\nu P \]
we have
\[ \big< \de L_{F}\,\big|  <\mathbb{P}_S\,|\, j^1 \varepsilon>\big> = 
-\frac{\D \mu}{\D
\rho}\; u^\cdot_\mu \left( \nabla_\nu J^\mu X^\nu - J^\nu
\nabla_\nu X^\mu +J^\mu
\nabla_\nu X^\nu \right)\, ds.\]
Integrating by parts, we get the first variation formula

\begin{equation}\begin{aligned} \big< \de L_{F}\,\big|  <\mathbb{P}_S\,|\, j^1 
\varepsilon>\big>=&
-\sqrt{\abs{g}}\left[(u^\cdot_\mu u^\nu + \de^\nu_\mu)  \nabla_\nu P + ( 
\mu + P )\: u^\nu \nabla_\nu
u^\cdot_\mu \right] X^\mu \,ds+\\
&- \frac{\mu +P}{\rho} u^\cdot_\al  X^\al (\nabla_\mu J^\mu)+\sqrt{\abs{g}}\nabla_\nu \left[ (\mu + P) (\de_\mu^\nu + u^\cdot_\mu 
u^\nu) X^\mu \right]\, ds+\\
&-\frac{\sqrt{\abs{g}}}{2} H^{(F)}_{\mu\nu} \de g^{\mu\nu}\,ds \end{aligned}\label{F}\end{equation}
with
\[\sqrt{\abs{g}}\,H^{(F)}_{\mu\nu}\, ds = 2\frac{\D L_F}{\D g^{\mu\nu}}= \sqrt{\abs{g}}\,[P g_{\mu\nu} -(\mu + P) u^\cdot_\mu u^\cdot_\nu ]\, ds.\]

Let us compute the first variation formula for the interaction term 
$L_{(int)}$:

\begin{equation}\begin{aligned} \big< \de L_{int}\,\big|  <\mathbb{P}_S\,|\, j^1 
\varepsilon>\big>&=\frac{\D
L_{int}}{\D J^\mu} \Lie_{X} J^\mu  + \frac{\D
L_{int}}{\D A_\nu}
\de A_\nu =\\
&= q A_\mu (\nabla_\nu J^\mu X^\nu - J^\nu \nabla_\nu X^\mu
+J^\mu \nabla_\nu X^\nu)\, ds + q J^\nu \de A_\nu ds=\\
&= q J^\nu \de A_\nu + q J^\mu F_{\mu\nu} X^\nu \, ds+ q \nabla_\nu
(A_\mu J^\mu  X^\nu - A_\mu J^\nu  X^\mu )\,ds
\end{aligned}\label{inter}\end{equation}

The complete set of the equations of motion for the Lagrangian 
\eqref{L1} is thence the following
\[\left\{\begin{aligned}{}&\nabla_\mu J^\mu=0\qquad\quad\text{(constraint)}\\
&R_{\mu\nu} -  \frac{1}{2} R g_{\mu\nu} = \kappa (H_{\mu\nu}^{(EM)} + 
H_{\mu\nu}^{(F)})\\
&\sqrt{\abs{g}}\nabla_\mu F^{\mu\nu} = q J^\nu\\
&(u^\cdot_\mu u^\nu
+ \de^\nu_\mu)  \nabla_\nu P + ( \mu + P )\: u^\nu \nabla_\nu 
u^\cdot_\mu - J^\mu
F_{\mu\nu}=0.
\end{aligned}\right.\]

We are now going to prove that the Lagrangian \eqref{L1} is variationally gauge-natural when restricted to the prolongation of the constraint submanifold $S\subset J^1 \mathfrak{C}(P)$ identified by equation $\nabla_\mu J^\mu=0$.

If $(x^\mu, \theta)$ are coordinates on the structure bundle $P$, an infinitesimal generator of principal automorphisms of $P$ is a vector field $ Y = Y^\mu (x^\al) \D_\mu + y(x^\al) \frac{\D}{\D \theta} \in \Gamma(TP)$
projecting onto $X= T\pi Y = Y^\mu \D_\mu$.
The Lie derivative of a section $\si$ of the gauge-natural configuration bundle  $ Lor(M) \times \mathcal{C(P)}  \times
\Lambda^3 T^\ast M$ with respect to $Y$ is
\[\Lie_Y \si = (\Lie_X g_{\al\B}) \frac{\D}{\D g_{\al\B}}+ (\Lie_Y A_{\nu})\frac{\D}{\D A_{\nu}} +  (\Lie_X
J^\mu) \frac{\D}{\D J^\mu}\]

with coefficients
\[\begin{aligned}
{}&\Lie_X g_{\al\B}= \nabla_\al Y^\cdot_\B + \nabla_\B Y^\cdot_\al\\
{}&\Lie_Y A_{\si} = Y^\rho F_{\rho\si}+ \nabla_\si (y + A_\rho Y^\rho)\\
{}&\Lie_Y F_{\al\B} = Y^\rho \nabla_\rho F_{\al\B} + \nabla_\al Y^\rho
F_{\rho\B}  +\nabla_\B Y^\rho  F_{\al\rho}\\
{}&\Lie_X J^\mu = \nabla_\nu J^\mu Y^\nu - J^\nu \nabla_\nu Y^\mu + J^\mu
\nabla_\nu Y^\nu.
\end{aligned}\]
Thus the configuration bundle is gauge-natural with order $(1,1)$.

A gauge-natural morphism $\mathbb{J}: J^{1} \mathfrak{C} \map (J^{1}\text{IGA}(P))^\ast \otimes_{J^1 \mathfrak{C}} \mathfrak{F}(P)$  such that we have \begin{equation}\big< \mathbb{P}_S\;|\; j^1 <\mathbb{J}|j^1 Y > \big>\circ \:j^1\si  = \Lie_Y \si \label{J}\end{equation} can be defined  by $<\mathbb{J}|j^1 Y>\circ \:j^1\si  = (\Lie_X g_{\al\B}, \Lie_Y A_{\si}, X)$.

\bigskip

Let us now consider the Lie derivative of the morphism $L$ with respect to $Y$. We find:

\[\begin{aligned}
\pi_2\Lie_Y L=
& \frac{\D L}{\D g_{\al\B}} \Lie_X g_{\al\B} + \frac{\D L}{\D \D_\la g_{\al\B}} \D_\la \Lie_X g_{\al\B} + \frac{\D L}{\D \D_{\la\omega} g_{\al\B}} \D_{\la\omega} \Lie_X g_{\al\B}+\\
& + \frac{\D L}{\D A_{\nu}} \Lie_Y A_{\nu} + \frac{\D L}{\D \D_\si A_{\nu}} \D_\si \Lie_Y A_{\nu} + \frac{\D L}{\D J^{\mu}} \Lie_X J^{\mu} -d_\mu (L Y^\mu).
\end{aligned}\]

Analyzing this term by term we have that the identities $\Lie_Y L_H=0$, $\Lie_Y L_{EM}=0$ are well known to be satisfied (see \cite{lo}) and  they can be recasted using the first variation formulae \eqref{H} and \eqref{EM} together with the identity \eqref{J} into 

\begin{equation}\frac{\sqrt{\abs{g}}}{2\kappa}(R_{\mu\nu} -  \frac{1}{2} R g_{\mu\nu}) \Lie_Y g^{\mu\nu}= \nabla_\al \mathcal{E}^\al_H \label{covH}\end{equation}
with
\[\mathcal{E}^\al_H ds_\al= \frac{\sqrt{\abs{g}}}{2\kappa}\left[\left(
\frac{3}{2}R^\al_{\cdot\la} -R\de^\al_\la\right) Y^\la+ 
\left( g^{\B\gamma}\de^\al_\la - g^{\al(\ga}\;\de^{\B)}_\la \right) \nabla_{\B\ga} Y^\la\right] \,ds_\al \]

and into

\begin{equation}- \sqrt{\abs{g}} (\nabla_\mu
F^{\mu\nu}) \Lie_Y A_\nu -\frac{\sqrt{\abs{g}}}{2}H^{(EM)}_{\mu\nu}\Lie_X g^{\mu\nu} = \nabla_\al \mathcal{E}^\al_{EM} \label{covEM}\end{equation}

with
\[\mathcal{E}^\nu_{EM}\, ds_\nu=
 -\sqrt{\abs{g}}\, \left[ g^{\nu\si}\,H^{(EM)}_{\si\mu}\,Y^\mu +F^{\nu\mu} \nabla_\mu ( y + A_\rho Y^\rho)\right] ds_\nu.\]

For the Lagrangian $L_F$ a straightforward calculation enables us to verify that the identity $\Lie_Y L_F=0$ holds too, and that it can be split (thanks to the first variation formula \eqref{F} together with identity \eqref{J}) into the following:

\begin{equation}\begin{aligned} {}&-\sqrt{\abs{g}}\left[(u^\cdot_\mu u^\nu + \de^\nu_\mu)  \nabla_\nu P + ( 
\mu + P )\: u^\nu \nabla_\nu
u^\cdot_\mu \right] X^\mu+\\
&- \frac{\mu +P}{\rho} u^\cdot_\al  X^\al (\nabla_\mu J^\mu) -\frac{\sqrt{\abs{g}}}{2} H^{(F)}_{\mu\nu} \Lie_X g^{\mu\nu} = \nabla_\al \mathcal{E}^\al_{F} \end{aligned}\label{covF}\end{equation}
with
\[\mathcal{E}^\al_F \, ds_\al= -\sqrt{\abs{g}}\,(g^{\al\mu}\, H^{(F)}_{\mu\nu}\, Y^\nu)\,ds_\al. \]

What remains is just to show that the Lagrangian $L_{int}$ is variationally gauge-natural when restricted to the constraint submanifold. In fact we have
\[\begin{aligned}
\pi_2 \Lie_Y L_{int} & =\left[\frac{\D L_{int}}{\D J^\mu} \Lie_X J^\mu + \frac{\D L_{int}}{\D A_\nu}
\Lie_Y A_\nu - \nabla_\al ( Y^\al L_{int})\right] \,ds =\\
&= q\,\big[ A_\mu (\nabla_\nu J^\mu Y^\nu - J^\nu \nabla_\nu Y^\mu
+J^\mu \nabla_\nu Y^\nu) + J^\nu [Y^\rho F_{\rho\nu}+ \nabla_\nu (y + A_\rho Y^\rho)]+\\
&\phantom{=}- A_\nu J^\nu \nabla_\al Y^\al-  Y^\al J^\nu \nabla_\al  A_\nu
-Y^\al  A_\nu \nabla_\al  J^\nu \big] \,ds  =\\
&=q\,\left[ J^\nu \nabla_\nu y \right] \,ds  = q\,\left[\nabla_\nu ( y \, J^\nu) - y \nabla_\nu J^\nu\right] \,ds 
\end{aligned}\]

where the first addendum is a (local) divergence, while the second one vanishes on the constraint.
Thanks to the first variation \eqref{inter} together with identity \eqref{J}, the previous formula can be recasted into the identity 
\begin{equation} q J^\nu \Lie_Y A_\nu + q J^\mu F_{\mu\nu} X^\nu  = \nabla_\al \mathcal{E}^\al_{int} \label{covint}\end{equation}
with
\[\mathcal{E}^\al_{int} = -q (y +A_\mu Y^\mu)J^\al\]
Summing up identities \eqref{covH}, \eqref{covEM}, \eqref{covF} and \eqref{covint} we find that the sum of the left hand sides vanishes if composed with a $\mathbb{P}_S$-critical section, while the current
\[\mathcal{E}^\al \, ds_\al= (\mathcal{E}^\al_{H}+\mathcal{E}^\al_{EM}+\mathcal{E}^\al_{F}+\mathcal{E}^\al_{int}) \, ds_\al\]
is conserved.

\bigskip

If in particular we compute the current $\mathcal{E}^\al \, ds_\al$ relative to the horizontal part $Y_{(\text{hor})}= Y^\mu \D_\mu - A_\rho Y^\rho \frac{\D}{\D \theta}$ of an infinitesimal generator of automorphisms of $P$, what we get is
\[\mathcal{E}^\al_{(\text{hor})} \, ds_\al = \mathcal{E}_{H}^\al \, ds_\al - \sqrt{\abs{g}}\,[g^{\al\mu}\,( H^{(F)}_{\mu\nu} + H^{(F)}_{\mu\nu})\, Y^\nu]\,ds_\al. \]
and this is the conserved current related with the energy-momentum of the system.

Moreover, the N\"other current relative to the vertical part $Y_{(V)}= (y +A_\mu Y^\mu) \frac{\D}{\D \theta}$ is 
\[\mathcal{E}^\al_{(V)} = -\sqrt{\abs{g}}\, F^{\al\mu} \nabla_\mu ( y + A_\rho Y^\rho) - (y +A_\mu Y^\mu)\,q J^\al.\]

Integrating by part the covariant derivative of $(y +A_\mu Y^\mu)$ we can split this vertical current according to

\[\mathcal{E}^\al_{(V)} =(\sqrt{\abs{g}} \nabla_\mu F^{\al\mu} -q J^\al) (y +A_\mu Y^\mu) -\nabla_\mu \big[ \sqrt{\abs{g}}\, F^{\al\mu} ( y + A_\rho Y^\rho)\big] \]

where the first summand vanishes on-shell while the argument of the divergence is called the {\it superpotential} $\mathcal{U}=-\frac{1}{2}\sqrt{\abs{g}} F^{\al\mu} ( y + A_\rho Y^\rho)ds_{\al\mu}$ and it is a $2$-form that is closed even off-shell.
We stress that the interaction Lagrangian does not contribute to the electromagnetic superpotential.

\section*{Conclusions}

Motivated by many physical and mathematical examples, we have studied variational problems with parametrized variations. In particular, starting from the results of \cite{fgr} we have worked on a twofold generalization: by one side we studied parametrized variational problems, no matter from where the parametrization comes from, and we have recovered and ever generalized known results. In a second moment we also focused on the application to vakonomic constraints, for which we have defined and compared the two different concepts of vak-criticality and $\mathbb{P}$-criticality.
A detailed study of nonholonomic field theory could also be based on the same framework and it shall certainly be performed in the future developments. 
Hopefully our unified language will also help to understand how constraints on the derivatives of fields have to be handled in the case of field theories.

For what concerns the second direction of generalization, we have formulated a variational theory for gauge-natural parametrized field theories including conserved currents that provides, in the case of relativistic hydrodynamics of a charged fluid, a non-conventional procedure to define field equations, N\"other currents and superpotentials and that turns out to be computationally much easier then the standard one.
A last observation that deserves to be further investigated is that a $\mathbb{P}$-Euler Lagrange equations arising from the parametrized variational problems is usually not variational in the standard sense. Nevertheless we have a very efficient machinery to study symmetries and conservation laws. Do this apply to any known non-variational equations?

\section*{Acknowledgments}
One of us (E.B.) dedicates this paper to his newborn daughter Lucia. We thanks prof. Raffaele Vitolo and prof. Marco Ferraris for very helpful discussions and personal support. We are also grateful to prof. Jerry Marsden and prof. Tudor Ratiu for a course held at the GNFM-INdAM Ravello summer school 2004 where some ideas began to grew. We acknowledge support from GNFM-INdAM, from MIUR Prin 2005 on \<<{\it Leggi di conservazione e termodinamica in meccanica dei continui e in teorie di campo}'' and from INFN-Iniziativa specifica NA12.

\appendix \section{Variational morphisms and their splittings}
\label{morfismi}
In this Appendix we report the theory of variational morphisms.
The theory is modeled on
\cite{lo} and \cite{fibered}, where complete proofs are given. 
To adapt
them to the constrained case we have to modify a bit the definitions. In
particular, the modified definitions of formal connection and 
variational morphism
seem to appear here for the first time.

\subsection{Variational morphisms}

Variational morphisms are an abstract model of the integrands 
appearing in global variational calculus.
They were introduced (see \cite{vm} for a discussion and some bibliography on alternative approaches) to provide a general framework in which we can 
implement two specific algorithmic procedures based on integration by parts. The first one explicitly construct a
splitting of such integrands into a volume and a boundary covariant part that
generalizes the first variation formula; the second provides superpotentials from the N\"other currents.

\DEF{Let $C\stackrel{\pi}{\map}M$ be a configuration bundle over a
$m$ dimensional base and
$E\stackrel{\pi_E}{\map}C$ a vector bundle over $C$. Let $k\geq h$ and 
$n$ be natural
numbers (zero included).

A {\it variational $E$-morphism} is a vertical (projecting onto the identity) bundle morphism
\[ \mathbb{M}: J^k C  \map (J^h E)^\ast {\otimes}_{J^h C}
\pi^\ast \Lambda^{m-n}T^\ast(M). \]
The minimal $k$ is called the {\it
order} of
$\mathbb{M}$, $h$ is the {\it rank}, while $m-n$ is the {\it degree} and $n$
the {\it codegree}.}

\subsection{Operations with (variational) morphisms}

\DEF{Given a vertical fibered morphism $\mathbb{F}=(\text{id}, \Phi)$ between 
two bundles $B$
and
$D$ over the same base $M$ we call $s$-order jet prolongation (or 
$s$-order formal
derivative) of $\Phi$ the unique fibered morphism $j^s \mathbb{F}=(\text{id},
j^s \Phi)$ that makes commutative the following diagram

\[
\begindc{\commdiag}[2]
\obj(20,10)[MB]{$M$}
\obj(20,40)[B]{$B$}
\obj(20,70)[JB]{$J^s B$}
\obj(50,10)[MD]{$M$}
\obj(50,40)[D]{$D$}
\obj(50,70)[JD]{$J^s D$}
\mor{B}{MB}{$\pi$}
\mor{JB}{B}{${\pi}^s_1$}
\mor{D}{MD}{$\pi'$}[\atright,\solidarrow]
\mor{JD}{D}{${\pi'}^s_1$}[\atright,\solidarrow]
\mor(20,9)(50,9){}[\atright,\solidline]
\mor(20,11)(50,11){id}[\atleft,\solidline]
\mor{B}{D}{$\Phi$}
\mor{JB}{JD}{$j^s\Phi$}
\cmor((19,14)(15,28)(19,37)) \pup(14,33){$\si$}[\solidline]
\cmor((19,14)(9,40)(19,67)) \pup(4,40){$j^s\si$}[\solidline]
\cmor((51,14)(55,28)(51,37)) \pup(61,35){$\Phi\circ\si$}[\solidline]
\cmor((51,14)(68,35)(51,67)) \pup(74,55){$j^s(\Phi\circ\si)$}[\solidline]
\cmor((16,36)(17,36)(19,37)) \pup(1,1){}[\solidline]
\cmor((18,34)(18,35)(19,37)) \pup(1,1){}[\solidline]
\cmor((18,64)(18,65)(19,67)) \pup(1,1){}[\solidline]
\cmor((16,65)(17,65)(19,67)) \pup(1,1){}[\solidline]

\cmor((54,36)(53,36)(51,37)) \pup(1,1){}[\solidline]
\cmor((52,34)(52,35)(51,37)) \pup(1,1){}[\solidline]
\cmor((52,64)(52,65)(51,67)) \pup(1,1){}[\solidline]
\cmor((54,66)(53,66)(51,67)) \pup(1,1){}[\solidline]
\enddc\]\label{prol}}

When one prolongs a morphism projecting onto the identity from a jet bundle to another bundle, e.g. $\mathbb{F}: J^p B\map D$, one 
gets a morphism
$j^1 \mathbb{F}: J^1(J^p B)\map J^1 D$. The bundle $J^{p+1} B$ is naturally 
embedded in
$J^1(J^p B)$ by a canonical inclusion $i$. We are often more interested into the map
$j^1 \mathbb{F} \circ i$ rather then in the prolongation itself. We call $j^1 \mathbb{F} 
\circ i$ the first order
{\it holonomic prolongation} of $\mathbb{F}$. In the sequel, when we mention the
prolongation of a morphism starting from a jet bundle we will always mean the
holonomic one.

\REMARK{We warn the reader that by $j^k$ we denote three different concepts of prolongation: prolongation of a section, of a vertical morphism and of a vector field. They are strongly linked, though not identical concepts.
For a discussion of this link we refer the reader to \cite{lo}, but we just remark that every vector field $X$ on a bundle $B$ can be seen as a vertical morphism $X:B\map TB$, and if we compute the first jet prolongation of this morphism we get another morphism $j^h X: J^h B \map J^h TB$ that is not a vector field on $J^h B$ as it should be if thought as the prolongation of a vector field.

If $(x^\mu, y^i_{\bar{\nu}})$ are coordinates on $J^h B$ ($\bar{\nu}$ is a multiindex of length $0\leq\abs{\bar{\nu}}\leq h$), and $(x^\mu, y^i_{\bar{\nu}}, v^\mu_{\bar{\nu}}, v^a_{\bar{\nu}})$ coordinates on $J^1 TB$, the coordinate expression of the prolonged morphism $j^h X$ is $j^h X(x^\mu, y^i_{\bar{\nu}}) =(x^\mu, y^i_{\bar{\nu}},d_{\bar{\nu}}v^\mu,d_{\bar{\nu}}v^a)$.

Luckily enough, for all integers $h$, a global canonical map $r_h: J^h TB \map  TJ^h B $ can be defined (see \cite{modugno, sardanashvily}), whose coordinate expression is $r_h(x^\mu, y^i_{\bar{\nu}}, v^\mu_{\bar{\nu}}, v^a_{\bar{\nu}})= (x^\mu, y^i_{\bar{\nu}},v^\mu, d_{\bar{\nu}} v^i - \sum_{\bar{\al}}y^i_{\bar{\al} +\bar{1}_\mu} d_{\bar{\nu}-\bar{\al}}v^\mu)$ where $\bar{\al}$ is any possible multiindex of length $0\leq\abs{\bar{\al}}<\abs{\bar{\nu}}=h$ and  $\bar{1}_\mu$ is a multiindex with $1$ entry in position $\mu$ and zero elsewhere. The ordinary prolongation $j^h X$ of the vector field $X$ as a vector field on $J^h B$ can be recovered just by composing $r_h$ with the prolonged morphism $j^X$.}

\DEF{Let $B\map M$ be a bundle and $E$ a vector bundle over the same 
base, let $\mathbb{M}$ be a fibered morphism $\mathbb{M}:  B\map
(E)^\ast $ and $\mathbb{P}$ another bundle morphism
$\mathbb{P}:  B \map  E$.

We define {\it formal contraction} \label{contraction}of $\mathbb{M}$ on
$\mathbb{P}$ the unique variational morphism
\[<\mathbb{M}\:|\: \mathbb{P} >   : B \map M\times\mathbb{R} \]
such that
\[\fa \rho \in \Gamma (B),\:\:
<\mathbb{M}\:|\: \mathbb{P}> \circ \rho = <\mathbb{M}\circ \rho \:
| \:\mathbb{P}\circ\rho>\] where $<\,|\,>$ in the right-hand side is the usual
fiberwise contraction between a section of a vector bundle and a section of its
dual. \label{contr}}

\DEF{\label{Div}
Let $\mathbb{M}: J^k C \map \Lambda^{m-n}T^\ast(M) $ be a variational morphism. We define the {\it divergence} of $M$ to be the unique variational morphism
$\rm{Div}(\mathbb{M}) :  J^{k+1} C \map \Lambda^{m-n+1}T^\ast(M) $ such
that for any section $\si \in
\Gamma (C)$ the following holds
\[\rm{Div}(\mathbb{M}) \circ j^{k+1} \si := d(\mathbb{M}\circ j^k \si).\]}

\subsection{Formal connections and fibered connections}
The following definitions generalize the ones given in 
\cite{fibered} to the
settings of constrained variational calculus.
Let us consider a composite projection $E\stackrel{\pi_E}{\map} B\stackrel{\pi}{\map} M$ and the fibered morphism (section)

\[
\begindc{\commdiag}[2]
\obj(20,10)[M]{$M$}
\obj(20,40)[B]{$B$}
\obj(50,40)[E]{$E$}
\mor{B}{M}{} 
\put {$\pi$} at 46 50
\mor{B}{E}{$\mathbb{F}$} 
\mor{E}{M}{} [\atright,\solidarrow]
\put {\mbox{\footnotesize$\pi\circ\pi_E$}} at 68 64
\cmor((19,14)(15,28)(19,37)) \pup(12,26){$\si$}[\solidline]
\cmor((16,36)(17,36)(19,37)) \pup(1,1){}[\solidline]
\cmor((18,34)(18,35)(19,37)) \pup(1,1){}[\solidline]
\cmor((25,12)(42,20)(50,36)) \pup(52,21){$\mathbb{F}\circ\si$}[\solidarrow]
\enddc \]

with $E$ vector bundle on $B$.

Let $(x^\mu, y^i)$ be a coordinate system on $B$ 
and $(x^\mu, y^i,
v^A)$ on
$E$ (with respect to a fiberwise local basis $\{e_A\}$) a local 
representation of
$\mathbb{F}$ is $v^A(x^\mu,y^i) e_A$, while a local 
representation of its first jet
prolongation (as a fibered morphism) $j^1 \mathbb{F}$ is
$v^A(x^\mu,y^i) e_A + d_\mu v^A e_A^\mu$, where
$\{e_A^\mu\}$ is a base of the fiber of $J^1 E\map E$ and
$d_\mu v^A = \D_\mu v^A + \D_i v^A y^i_\mu$ is called the {\it formal 
derivative}
of $\mathbb{F}$.

The transformation rules  of the formal derivatives under a fibered change of
coordinates are not tensorial.

Consider a set
$\{\Gamma^A_{B \mu}(x^\mu,y^i, y^i_\mu)\}_{ 
A,B\in \{1, \cdots, \text{rank} E\},\; 
\mu\in\{1,\cdots , \dim M\} }$
of local coefficients that fit into the expression
\begin{equation}\nabla_\mu v^A =d_\mu v^A + \Gamma^A_{B \mu
}v^B\label{fcd}\end{equation} giving rise to an object at the left 
hand side of the
equalities that under a change of fibered coordinates	
\[\begin{array}{ccc}\left\{\begin{aligned}
{}&x^{\mu'}=x^{\mu'}(x^\mu)\\
&y^{i'}=y^{i'}(x^\mu, y^{i})\\
&y^{i'}_{\mu'}= (J^{-1})^\mu_{\mu'} d_\mu y^{i'}=(J^{-1})^\mu_{\mu'} 
(J^{i'}_\mu
+J^{i'}_i y^i_\mu) \\
&v^{A'}=M^{A'}_A (x^{\mu}, y^{i}) v^A 
\end{aligned}\right.&\qquad&\begin{aligned}
{}&J^{\mu'}_\mu=\frac{\D x^{\mu'}}{\D x^\mu}\\
&J^{i'}_\mu=\frac{\D y^{i'}}{\D x^\mu}, \quad J^{i'}_i=\frac{\D 
y^{i'}}{\D y^i}\\
&M^{A'}_A\in GL(\text{rank} E, \R)\end{aligned} \end{array}\]
transforms according to the rule
\begin{equation}\label{1A}\nabla_{\mu'} 
v^{A'}=(J^{-1})^\mu_{\mu'}\nabla_\mu v^A
M^{A'}_A\end{equation} that are nothing but
the transition functions of the vector bundle $E \otimes_B T^\ast M\map B$.

The resulting transformation law for the coefficients is
\begin{equation} \label{1B}\Gamma^{A'}_{B'{\mu'}} = 
J^\mu_{\mu'}(\Gamma^A_{B \mu}
M_A^{A'} M^B_{B'}  - d_\mu M^{A'}_A M_{B'}^B)\end{equation} It's easy 
to verify that
relations
\eqref{1B} form a cocycle over the manifold
$J^1B$, thus they define the transition functions of an affine bundle 
$FC(E)\map
J^1B$ modeled on the vector bundle $E^\ast \otimes_B E \otimes_B T^\ast M$.

\DEF{A {\it formal connection } on the composite fiber bundle $E\map B\map M$  is a global section of the affine 
bundle $FC(E)$.\label{formalcon}}
Let us consider the bundle $\mathbb{L}(E)\map B$ of linear frames of the vector bundle
$E\stackrel{\pi_E}{\map} B$. Let $r=\text{rank} E$.
$\mathbb{L}(E)$ is a principal $GL(r, \R)$ bundle and the free natural
right action $r: \mathbb{L}(E) \times GL(r, \R) \map \mathbb{L}(E)$ can be defined on it.

Let us call $\mathbb{L}_M(E)$ the bundle $\mathbb{L}(E)\map M$ obtained by the 
obvious composition
of projections. It is no longer principal, nevertheless the 
action $a$ induces
  on it a free right action $a_M$, that can be lifted to a right 
action $j^1 a_M $
on its fist jet extension $J^1 \mathbb{L}_M(E)$. The action $j^1 a_M 
$ is still free and
admits a quotient manifold $\frac{J^1 \mathbb{L}_M(E)}{GL(r, \R)}$ 
that has an affine
bundle structure over the base $J^1B$. The two affine bundles $FC(E)$ and
$\frac{J^1 \mathbb{L}_M(E)}{GL(r, \R)}$ are isomorphic and this can be proved directly by 
showing that they
share the same transition functions, providing a more
intrinsic characterization of the bundle of formal connections.

Consider now the bundle $J^1 E_{J^1B} \stackrel{j^1 \pi_E}{\map} J^1 B$; the
presence of a formal connection on $E$, allows us to define the
global vector bundle morphism $\nabla$ by the following commutative diagram

\[
\begindc{\commdiag}[2]
\obj(20,10)[JB]{$J^1B$}
\obj(20,40)[JE]{$J^1 E_{J^1B}$}
\obj(55,10)[B]{$B$}
\obj(55,40)[E]{$E \otimes_B T^\ast M$}
\mor{JE}{E}{$\nabla$} 
\mor{JB}{B}{${\pi_B}^1_0$} 
\mor{JE}{JB}{} 
\mor{E}{B}{}
\enddc\]

whose local coordinate expression is \eqref{fcd}.

\DEF{Let $\si$ be a section of $J^1 E_{J^1B}$. We call $\nabla 
\circ\si$ the
{\it formal covariant derivative} of $\si$. }

\DEF{A {\it fibered connection} on $E\stackrel{\pi_E}{\map}B$ is a 
couple $(\gamma,
\Gamma)$, where $\gamma$ is a linear connection on $M$ and $\Gamma$ 
is a formal
connection on $E$.\label{fibercon}}

Given a fibered connection on $E$ we can extend the morphism $\nabla$ to $J^1
(E_{J^1B} \otimes T^p_q(M))$ where $T^p_q(M)$ is the algebra of 
$p$-times covariant
and
$q$-times contravariant tensors on $M$ with the obvious tensorization 
procedure,
getting a fibered morphism

\[
\begindc{\commdiag}[2]
\obj(20,10)[JB]{$J^1B$}
\obj(20,40)[JE]{$J^1 (E_{J^1B}\otimes T^p_q(M))$}
\obj(75,10)[B]{$B$}
\obj(75,40)[E]{$E \otimes_B T^p_{q+1}(M)$}
\mor{JE}{E}{$\nabla$} 
\mor{JB}{B}{${\pi_B}^1_0$} 
\mor{JE}{JB}{} 
\mor{E}{B}{}
\enddc\]
that we denote by the same symbol $\nabla$.

\DEF{Let $\si$ be a section of $J^1 (E_{J^1B}\otimes T^p_q(M))$, let 
us call $\nabla
\circ\si$ the {\it formal covariant derivative} of $\si$. }

\DEF{The {\it formal covariant derivative} of a morphism 
$\mathbb{F}:B \map E$ is
the unique fibered morphism $\nabla \mathbb{F}: J^1 B \map E 
\otimes_B T^\ast M$ such
that
\[\fa \rho\in \Gamma (B), \nabla \mathbb{F} \circ j^1\rho = \nabla\circ
j^1 (\mathbb{F} \circ \rho).\]}

A formal connection on a composite fiber bundle $E\map B\map M$ can be constructed from a linear connection $A$ on the vector bundle $E\map B$ and the first jet prolongation $j^1 \si$ of a section of the bundle $B\map M$.
Let $(x^\mu, y^a, v^B)$ be a fibered coordinate system on $E$; the connection $A$ has the coordinate representation \[A= dx^\mu \otimes (\frac{\D}{\D x^\mu} - A^B_{D \mu} v^D \frac{\D}{\D v^B}) + dy^i \otimes (\frac{\D}{\D y^i} - A^B_{D i} v^D \frac{\D}{\D v^B}).\]

The coefficients $ \Gamma^B_{D \mu}= A^B_{D \mu} + A^B_{D i} y^i_\mu$ transform according to \eqref{1B} and thus they glue together into a global formal connection on $E\map B \map M$.

\subsection{Local expression and reduction of a variational morphism with
respect to a fibered connection}
Fixed a fiberwise basis $\left\{e_A (x),e_A^{\la_1}(x),\cdots , e_A^{{\la_1} \cdots
{\la_h}}(x)\right\}_{ A\in
\{1, \cdots, p\},\;\; i \in \{1, \cdots, h\}, \;\; \la_i \in \{ 1, \cdots,  m
\} }$ of $J^h_x E$ and its dual
  $\left\{e^A
(x),e^A_{\la_1}(x),\cdots , e^A_{{\la_1} \cdots {\la_h}}(x)\right\}$ in $(J^1_x
E)^\ast$, the local expression of a variational morphism
$\mathbb{M}: J^k C   \map (J^h E)^\ast \otimes\Lambda^{m-n}T^\ast(M)$ is the
following

\begin{equation}
\mathbb{M} = \frac{1}{n!}\left( v^{{\mu_1}\cdots  {\mu_n}}_A e^A
+v^{{\mu_1}\cdots {\mu_n}{\la_1}}_A e^A_{\la_1}+ \cdots + v^{{\mu_1}\cdots
{\mu_n}{\la_1} \cdots {\la_h}}_A   e^A_{{\la_1} \cdots {\la_h}}\right)
\otimes ds_{{\mu_1}\cdots {\mu_n}},\label{local1}
\end{equation}

where the indices ${\mu_1}\cdots {\mu_n}$ are  skew-symmetric while the
${\la_1} \cdots {\la_h}$ (if any) are  symmetric and where the
coefficients are functions of the point  $j_x^k \rho \in J^k C$.

\bigskip

Given a fibered connection $(\gamma^\al_{\B\mu}, \Gamma^A_{B\mu} )$ 
we can change
basis on  $J_x^h E$ in a way that a holonomic section $j^h X\in 
\Gamma (J^h E)$ in
the new basis  (called {\it basis of symmetrized covariant derivatives}) has
components

\[\begin{aligned}
{}&\hat{v}^A = v^A (x)\\
&\hat{v}^A_{\la_1} = \nabla_{\la_1}  v^A (x)= d_{\la_1} v^A  +
\Gamma^A_{B{\la_1}} v^B\\ &\vdots\\
&\hat{v}^A_{{\la_1}\cdots {\la_h}} =  \nabla_{(\la_1} \cdots
\nabla_{{\la_n})} v^A (x).\end{aligned}\]

Accordingly, using the dual basis on $(J^h E)^\ast$ we
can define a new coordinate system $(x^\al,
\hat{v}_A^{{\mu_1}\cdots {\mu_n}},  \hat{v}_A^{{\mu_1}\cdots
{\mu_n}{\la_1}}, \hat{v}_A^{{\mu_1}\cdots {\mu_n}{\la_1}\cdots {\la_h}})$
on $(J^h E)^\ast \otimes_{J^hC}\Lambda^{m-n}T^\ast(M)$ such that the local 
expression of
$ \mathbb{M}$ is
\[
\mathbb{M} = \frac{1}{n!} \left( \hat{v}^{{\mu_1}\cdots {\mu_n}}_A
\hat{e}^A  + \hat{v}^{{\mu_1}\cdots {\mu_n}{\la_1}}_A
\hat{e}^A_{\la_1}+ \cdots +  \hat{v}^{{\mu_1}\cdots {\mu_n}{\la_1} \cdots
{\la_h}}_A  \hat{e}^A_{{\la_1} \cdots {\la_h}}\right) \otimes
ds_{{\mu_1}\cdots {\mu_n}}.
\]
We remark that in the previous expression,  as in formula \eqref{local1},
the coefficients  have to be meant as functions of the point $j_x^k
\rho \in J^k C$, and the indices ${\mu_1}\cdots {\mu_n}$ are
skew-symmetric while the ${\la_1}
\cdots {\la_h}$ (if any) are symmetric.

The explicit change of basis and coordinates in the case $h=1$ is
\[\begin{array}{ccc} \left\{\begin{aligned}{}&\hat{e}^A = e^A\\
&\hat{e}^A_\la = e^A_\la + \Gamma^{A}_{B\la} e^B\end{aligned}\right.&
\text{and} &\left\{\begin{aligned}{}&v_A= \hat{v}_A  + \hat{v}_B^\la
\Gamma^{B}_{A\la}\\ &v_A^\la = \hat{v}_A^\la\end{aligned}\right.\end{array}\]

The introduction of a fibered connection can  select a \<<preferred'' kind
of variational  morphism having nice properties when written in the
symmetrized covariant derivatives coordinate system.

\DEF{\label{redu}Let $ \mathbb{M}: J^k C  \map  (J^h E)^\ast
\otimes_{J^hC}\Lambda^{m-n}T^\ast(M)$ be a variational morphism and
$\mathbb{M}_l$ its term of rank $0\leq l \leq h$. The term $\mathbb{M}_l$
is said to be {\it reduced} with respect to the fibered connection
$(\gamma^\al_{\B\mu}, \Gamma^A_{B\mu} )$ if, written in the symmetrized
covariant derivatives coordinate system, it has the coefficients symmetric in
the first $n+1$ upper indices, i. e. if $\hat{v}^{[{\mu_1}\cdots
{\mu_n}{\la_1}] \cdots {\la_l}}_A=0$. The variational morphism
$\mathbb{M}$  is reduced if each one of its terms is reduced.}

Notice that the term of rank zero is always reduced  with respect to any
connection,  having always antisymmetric upper indices and that in the
case of Mechanics ($m=1$) all variational morphisms are reduced with
respect to any connection.

Now we are ready to state the most important results of the  theory, that
justify the name  \<<variational'' for the morphism we have introduced,
providing a sort of \<<first variation formula'' from any variational
morphism. The proofs of the next two Theorems can be found in 
\cite{lo}, but as a
consequence of the minor modifications we introduced here one has to
reinterpret the symbols according to the previous definitions.

\THEO{{\bf-\<<Splitting lemma''-}\label{splitting}
Let $\mathbb{M}: J^k C  \map (J^h E)^\ast
\otimes_{J^hC}\Lambda^{m }T^\ast M$  be a global variational morphism with
codegree $n=0$ and let $h$, $k$ $\in \N$ (zero included). Chosen a fibered
connection $(\ga, \Gamma)$ on $E$ there exist a unique pair $\mathbb{V},
\mathbb{B}$ of variational morphisms
\[\begin{aligned}
{}&\mathbb{V}\equiv \mathbb{V}(\mathbb{M})\: : J^{h+k} C   \map
E^\ast \otimes_C\Lambda^{m}T^\ast M\\ &\mathbb{B}\equiv
\mathbb{B}(\mathbb{M},\ga)\:  : J^{h+k-1} C   \map (J^{h-1}
E)^\ast
\otimes_{J^{h-1}C}\Lambda^{m-1}T^\ast M
\end{aligned}\]
reduced with respect to $(\ga, \Gamma)$ such that $\fa V \in \Gamma(E)$ the
following holds true
\begin{equation}
<\mathbb{M}\:|\:j^h V > = <\mathbb{V}\:|\: V > + \text{ Div}<\mathbb{B}
\:|\:j^{h-1} V >.\end{equation} The variational morphism $\mathbb{V}$ is
called the volume part of $\mathbb{M}$ and $\mathbb{B}$  its boundary
part.}

\THEO{{\bf-\<<Reduction lemma''-} Let $\mathbb{M}: J^k C   \map (J^h
E)^\ast \otimes_{J^h C}\Lambda^{m-n}T^\ast M$  be a global variational
morphism of codegree $n\geq 1$ and let $h$, $k$ $\in \N$ (zero included). 
Chosen a fibered
connection $(\ga, \Gamma)$ on $E$ there exist a unique pair $\mathbb{V},
\mathbb{B}$ of variational morphisms
\[\begin{aligned}
{}&\mathbb{V}\equiv \mathbb{V}(\mathbb{M})\: : J^{h+k} C   \map
(J^h E)^\ast
\otimes_{J^hC}\Lambda^{m-n}T^\ast M\\ &\mathbb{B}\equiv \mathbb{B}(\mathbb{M},\ga)\:
: J^{h+k-1} C  \map (J^{h-1}
E)^\ast \otimes_{J^{h-1}}
\Lambda^{m-n-1}T^\ast M
\end{aligned}\]
reduced with respect to $(\ga, \Gamma)$ such that $\fa V \in \Gamma(E)$ the
following holds true
\begin{equation}
<\mathbb{M}\:|\:j^h V > = <\mathbb{V}\:|\:j^h V > + \text{ Div} <\mathbb{B}\:|\:j^{h-1} V >.\end{equation} 
The variational
morphism $\mathbb{V}$ is called the volume part of $\mathbb{M}$ and
$\mathbb{B}$ its boundary part.
\label{reduction}}

Constructive proofs of these Theorems are given in \cite{lo} where we can find a collection of older results; they are carried on by induction on the rank of the
morphism. If the rank is $1$ then one just needs to integrate by parts and no fibered connection is needed. At any rank they provide an algorithm out of which we can perform explicitly the 
splitting or the reduction.

In both the Theorems the volume part is uniquely defined even if we do
not require it to be reduced, and finally it does not depend on the fibered 
connection.
The boundary parts, on the contrary, are defined modulo a divergenceless term, but the
condition of being reduced with respect to any specific fixed fibered connection 
determines them uniquely (if the rank is $1$ the boundary part is stiill unique and reduced with respect to any fibered connection).

Moreover it was proved \cite{kolarticolo} that the boundary parts depend 
in fact only
on the connection on the base, while the formal connection on $E$ 
reveals itself to be just an
intermediate object useful to explicitly construct the splitting, but 
in the end unessential.
\section{Gauge natural bundles}

\label{GNmaterial}

\DEF{Let $G$ be a Lie group. $\mathfrak{P}^n(G)$ will denote the category of principal $G$-bundles whose base manifold is $n$-dimensional and whose maps are G-bundle maps projecting onto local diffeomorphism between the two bases.}
\DEF{ A covariant functor $\mathfrak{B}$ from the category $\mathfrak{P}^n(G)$ to the category of bundles and bundle morphisms is a {\it gauge-natural functor} if
\begin{enumerate}
\item[a)] for all principal $G$-bundle $P\stackrel{p}{\map}M$, 
$\mathfrak{B}(P)$ is
a bundle with structure group $G$ having the same base $M$ as $P$;
\item[b)] any principal morphism $\Phi: P\map P'$ projecting onto 
$\phi: M\map M$
induces a fibered morphism $\mathfrak{B}(\Phi) : \mathfrak{B}(P) \map
\mathfrak{B}(P')$ also projecting over $\phi$;
\item[c)] for any open $U\subseteq M$ the inclusion
morphism $i: p^{-1}(U) \map P$ is mapped into the inclusion morphism 
$\mathfrak{B}(i):
\mathfrak{B}(p^{-1}(U)) \map\mathfrak{B}(P)$.
\end{enumerate}}

\DEF{A {\it gauge-natural bundle} $GN \stackrel{\ga}{\map} M$ having 
the principal
bundle $P\stackrel{p}{\map}M$ as {\it structure bundle} is the image of $P$ trough a gauge-natural
functor $\mathfrak{B}$.}
Examples of gauge-natural bundles are all natural  bundles (trivial 
case), the bundle of
principal connections $\mathcal{C}(P)$ on a principle bundle $P$, the bundle
$\text{IGA}(P)$ of its Infinitesimal Generators of principal Automorphisms, the  bundle of
spin frames and others (see \cite{lo}).

Notice that jet prolongations of a gauge-natural bundle are 
gauge-natural with the same structure
bundle, and that every composite fibration 
$Z\stackrel{\pi_Z}{\map}Y\stackrel{\pi_Y}{\map}X$
where $Y$ is a gauge-natural bundle with structure bundle $P$ and 
$\pi_Z$ is natural makes
$\pi_Y\circ\pi_Z$ a gauge-natural bundle with structure bundle $P$.

\bigskip

We want to stress that for our purposes, the most relevant property 
of a gauge-natural bundle $\mathfrak{B}(P)$ is the possibility of lifting local
principal automorphisms of the structure $G$-bundle $P$ to local fibered morphisms of $\mathfrak{B}(P)$. We will call \<<{\it gauge transformation}'' on $\mathfrak{B}(P)$ the gauge-natural lift of a local principal automorphism of $P$ and
\<<{\it gauge group}'' the structure group $G$.

The same, via the tangent map, can be done
for infinitesimal generators: an infinitesimal generator of principal automorphisms can be
lifted to a vector field on $\mathfrak{B}(P)$ that will be called its \<<{\it gauge-natural lift}''.

Let us consider a $1$-parameter family of infinitesimal generators of principal automorphisms $\{\Psi_s= (\phi_s, \psi_s)\}$.
It can be proved (see \cite{lo}) that any infinitesimal generator $\Xi\in \Gamma(TG)= \frac{d}{ds}\Psi_s|_{s=0}$ is a right invariant projectable vector field on $P$ and that we can explicitly construct the bundle $\text{IGA}(P)\map M$ whose sections are in one to one correspondence with the infinitesimal generators of such families and with right invariant sections of $TP$. 
Infinitesimal generators of vertical principal automorphisms $\{\Psi_s= (id, \psi_s)\}$ can be identified with sections of a subbundle $\textstyle\frac{VP}{G}\hookrightarrow \text{IGA}(P)$ that is associated to $P\stackrel{p}{\map}M$ by means of the adjoint action of the Lie algebra $\mathfrak{g}$ of the gauge group $G$.

By means of a trivialization of $P$ we can locally define a set $\{\rho_A\}_{A\in\{1\cdots \dim G\}}$ of right invariant vector fields on $P$ that at every point form a basis of $VP$.
Let $(x^\mu, y^a)$ be a fibered system of coordinates on $P$; every infinitesimal generator of principal automorphisms $\Xi$ can be locally written as a right invariant vector field 
\[\Xi (x^\mu, y^a)= \xi^\mu (x^\mu) \frac{\D}{\D x^\mu} + \xi^A (x^\mu)\rho_A(x^\mu, y^a)\]
on $P$.

\DEF{We say that a gauge-natural bundle $\mathfrak{B}(P)$  is of $(r,
s)$-order ($s\leq r$) if the gauge-natural lift is a morphism
\[ \mathbb{L}:\mathfrak{B}(P) \map (J^r \text{IGA(P)})^\ast \otimes_{\mathfrak{B}} T\mathfrak{B}(P) \]
such that if $\Xi$ is vertical (i.e. it is a section of $\textstyle\frac{VP}{G}\hookrightarrow \text{IGA}(P)$), the vector field $\hat{\Xi}_\mathfrak{B} = <\mathbb{L}\;|\; j^r\Xi>$ on $\mathfrak{B}(P)$ is vertical, too, and depends in fact on the derivatives of $\Xi$ only up to the order $s$.}

\DEF{
Let $\Psi= (\phi, \psi)$ be a local principal automorphism of $P$ and 
$\mathfrak{B}(\Psi)=(\phi, \hat{\Psi}_\mathfrak{B})$ be its gauge-natural lift to the gauge-natural bundle
$\mathfrak{B}(P)$. Let $\si$ be a section of $\mathfrak{B}(P)$; we define its {\it push-forward} along $\Psi$ to be the section $\Psi^\star \si = \hat{\psi}_\mathfrak{B} \circ \si \circ \phi^{-1}$.}

\DEF{Let $\Xi$ be the infinitesimal generator of the $1$-parameter family of local principal automorphisms $\{\Psi_s= (\phi_s, \psi_s)\}$; we define the {\it Lie derivative} of a section $\si$ of $\mathfrak{B}(P)$ along $\Xi$ to be the unique section of the pull back bundle $\Lie_\Xi \si\in \Gamma (\si^\star V\mathfrak{B}(P))$ such that
\[ \Lie_\Xi \si = - \frac{d}{ds} \Psi_s^\star \si. \]}
\REMARK{\label{dlm}We can easily prove that if $\hat{\Xi}_\mathfrak{B}$ is the gauge-natural lift of $\Xi$ on $\mathfrak{B}(P)$, and $\Xi$ is the projection of $\Xi$ on $M$, the Lie derivative of a section can be computed trough the following formula
\[\Lie_\Xi \si = T\si \circ \pi \Xi - \hat{\Xi}_\mathfrak{B}.\]
In coordinates, if $(x^\mu,y^i)$ are fibered coordinates on $\mathfrak{B}(P)$ while the lift can be written as $\hat{\Xi}_\mathfrak{B}= \xi^\mu \frac{\D}{x^\mu}+ \hat{\Xi}_\mathfrak{B}^i \frac{\D}{\D y^i}$ one has $\Lie_\Xi \si=\Lie_\Xi y^i \frac{\D}{\D y^i}$, with
\[ \Lie_\Xi y^i= d_\mu y^i(x)  \xi^\mu (x) - \hat{\Xi}_\mathfrak{B}^i(x, y(x)) \]

\bigskip

It also turns out that the Lie derivative operator on sections of the gauge-natural bundle $\mathfrak{B}(P)$ of order $(r,s)$ (where $r\geq s$) can be interpreted as a fibered morphism
\[ \mathbb{D}_\Lie : J^1 \mathfrak{B}(P) \map (J^r \text{IGA(P)})^\ast \otimes V\mathfrak{B}(P)\]
such that for all sections $\si\in\Gamma(\mathfrak{B}(P))$ and for all infinitesimal generator of principal automorphism $\Xi\in \Gamma (\text{IGA}(P))$ one has $< \mathbb{D}_\Lie \;|\; J^r \Xi> \circ\; j^1 \si = \Lie_\Xi \si$. }
Notice that if $\Xi$ is vertical (i.e. it is a section of $\textstyle\frac{VP}{G}\hookrightarrow \text{IGA}(P)$) then $\Lie_\Xi \si$ depends in fact on the derivatives of $\Xi$ only up to the order $s$. 

\REMARK{\label{pi2}When in differential geometry one computes the Lie derivative of a vector field on a manifold $M$ one gets another vector field and so happens for differential forms and tensors. According to our definition, on the contrary, the Lie derivative of a vector field is a section of the vertical bundle $VTM$. To reconcile this apparent contradiction we have to remark that whenever $\mathfrak{B}$ is a vector bundle with base $M$ then  its vertical $V\mathfrak{B}$ is isomorphic (see \cite{kolar}) to the fibered product $\mathfrak{B}\times_M \mathfrak{B}$ and one has a global vertical fibered morphism $\pi_2:V\mathfrak{B}\map \mathfrak{B}$ that realizes the projection onto the second fibered factor.
If $(x^\mu, y^a, v^a)$ is a local coordinate system on $V\mathfrak{B}$ then the map $\pi_2$ acts as follows:
$\pi_2:(x^\mu, y^a, v^a)\associa (x^\mu, v^a).$
If $\si$ is a section of a gauge-natural vector bundle then $\pi_2\Lie_\Xi \si$ would be what traditionally is meant as its Lie derivative, the coordinate computation being identical, but with a slightly different interpretation.
}

\bigskip

Let $P\stackrel{p}{\map}M$ be a principal bundle; let moreover $U\subset M$ 
be open, $(x^\mu, g^a)$ be a coordinate system on $p^{-1}(U)\subset P$ and $\{\rho_A   (x^\mu,
g^a)\}_{A=1\cdots
\dim G}$  a fiberwise right invariant basis of vertical vector 
fields. An infinitesimal
generator of principal automorphisms has the local form $\Xi(x^\mu, 
g^a)= \xi(x)^\mu \D_\mu
+\xi^A(x) \rho_A (x^\mu, g^a)$ (see \cite{lo}). Given a 
principal connection $\om
(x^\mu, g^a)= dx^\mu \otimes (\D_\mu -\om^A_\mu (x) \rho_A(x^\mu, 
g^a))$  we can also split $\Xi$ into a vertical and a horizontal part according to
\[\Xi= \xi^\mu (\D_\mu -\om^A_\mu \rho_A) \oplus (\om^A_\mu \xi^\mu  + 
\xi^A) \rho_A =
\Xi_{(H)}\oplus \Xi_{(V)}.\]

\bibliographystyle{plain}
\bibliography{bibliografia}
\end{document}